\documentclass[a4paper,10pt,notitlepage]{article}
\usepackage{graphicx}
\usepackage{amsfonts,amssymb,amsmath,amsthm}
\usepackage{bbm}
\usepackage{mathrsfs}
\usepackage{url}
\usepackage{cite}
\usepackage{slashed}
\usepackage{psfrag}
\usepackage{bm}
\usepackage{multirow}

\numberwithin{equation}{section}

\newcommand\id{\mathbbm{1}}

\newenvironment{assumption}[2][Assumption]{\begin{trivlist}
\item[\hskip \labelsep {\bfseries #1}\hskip \labelsep {\bfseries #2}]}{\end{trivlist}}

\newtheorem{claim}{Claim}
\DeclareMathOperator{\wt}{wt}

\DeclareMathOperator{\Sym}{Sym}
\DeclareMathOperator{\spec}{spec}

\DeclareMathOperator{\osp}{osp}
\DeclareMathOperator{\so}{so}
\DeclareMathOperator{\un}{u}
\DeclareMathOperator{\gl}{gl}
\DeclareMathOperator{\sgl}{sl}

\DeclareMathOperator{\str}{str}

\DeclareMathOperator{\Img}{Im}

\DeclareMathOperator{\End}{End}

\title{\textbf{Continuum Limit of gl$\mathbf{(M|N)}$ Spin Chains}}

\author{Constantin Candu
\bigskip \\DESY
Hamburg, Theory Group, \\
Notkestrasse 85, D--22607 Hamburg, Germany
\bigskip\\ e-mail: \url{constantin.candu@desy.de}
}
\date{}

\begin{document}
\begin{titlepage}
\maketitle
\begin{abstract}
We study the spectrum of an integrable anti\-ferromagnetic Hamiltonian of the $\gl(M|N)$ spin chain of alternating fundamental and dual representations.
After extensive numerical analysis, we identify the vacuum and low lying excitations and with this knowledge perform the continuum limit, while keeping a finite gap.
All $\gl(n+N|N)$ spin chains with $n,N>0$ are shown to possess in the continuum limit $2n-2$ multiplets of massive particles which scatter with $\gl(n)$ Gross-Neveu like $S$-matrices, namely their eigenvalues do not depend on $N$. We argue that the continuum theory is the $\gl(M|N)$ Gross-Neveu model.
We then look for remaining particles in the $\gl(2m|1)$ chains.
The results suggest there is a continuum of such particles, which in order to be fully understood require  finite volume calculations.

\end{abstract}
%\vspace*{-13cm} {\tt {DESY xxx}}
%
%\hfill {\tt {yymm.nnnn}}
%\\
%{\tt {xxx}} \hfill {\tt {xxx}}
%\bigskip
\end{titlepage}

\tableofcontents

\addtolength{\baselineskip}{3pt}

%%%%%%%%%%%%%%%%%%%%%%%%%%%%%%%%%%%%%%
%%%%%%%%%%%%%%%%%%%%%%%%%%%%%%%%%%%%%%

\section{Introduction}

%%%%%%%%%%%%%%%%%%%%%%%%%%%%%%%%%%%%%%
%%%%%%%%%%%%%%%%%%%%%%%%%%%%%%%%%%%%%%

At present, the continuum limit of antiferromagnetic Lie superalgebra spin chains is poorly understood. High degeneracies or continuous spectra~\cite{Essler:2005ag, Saleur:2006tf, 2006cond.mat.12037I} and singular IR scattering amplitude behavior~\cite{Essler:2005ag, Saleur:2009bf} emerge in the continuum large volume limit. We interpret this as a hint of new physical phenomena.
Strikingly, when $q$-deformed, connections with long known puzzling theories with no obvious supergroup symmetry appear~\cite{Saleur:2009bf}.

Previous, above mentioned, research has mostly dealt with  $\osp(2|2)\simeq \sgl(2|1)$ spin chains.
We adopt a general approach, taking as a model~\cite{Essler:2005ag}, and hoping that the greater $\gl(M|N)$ symmetry and the relatively new cohomology techniques developed in~\cite{Candu:2010yg} will allow us to better understand the nature of these new phenomena, thus, suggesting how to correctly interpret, treat and, eventually, apply them.
The goal is to ultimately formulate a meaningful factorizable scattering theory for the continuum limit of superspin chains.
This is an even harder task then it sounds, because there are few massive relativistic field theories with supergroup symmetry that have been understood, at least partially~\cite{Saleur:2001cw, Saleur:2009bf, Bassi:1999ua, 2000NuPhB.583..475G}.

So, clarifying the continuum limit of superspin chains must ultimately result in a better understanding of formally simple field theories such as the $\gl(M|N)$ and $\osp(R|2S)$ Gross-Neveu models, which play an important role in disordered electronic system~\cite{Bernard:1995as, 2000NuPhB.583..475G, LeClair:2007aj}, provide instances of continuous families of conformal field theories~\cite{Candu:2008yw, Candu:2009ep} and appear in strong-weak coupling dualities with sigma models~\cite{Candu:2008yw, 2000NuPhB.583..475G, Creutzig:2008an}.
Due to strong violation of unitarity, it is not obvious how to treat these super Gross-Neveu models directly in the continuum by standard bootstrap methods~\cite{Zamolodchikov:1978xm, ORW}.
Insight from coordinate Bethe ansatz for chiral Gross-Neveu models~\cite{Andrei:1979un, Andrei:1979wy, Andrei:1979sq, Destri:1987ug} has led us to consider the spin chain of $\gl(M|N)$ alternating fundamental and dual representations as the most natural candidate for an integrable discretization of the $\gl(M|N)$ Gross-Neveu model.
The spin chain of $\gl(M|N)$ fundamental representations is discarded because it cannot lead to a relativistic field theory~\cite{Saleur:1999cx, 1994IJMPB...8.3243E}.
This can be immediately seen from the fact that the dual of any $\gl(M|N)$ representation appearing in the tensor product of $\gl(M|N)$ fundamental representations does not belong to this tensor product if $N>0$.

The article is organized as follows.
In sec.~\ref{sec:formalism} we define the spin chain and its integrable dynamics.
In sec.~\ref{sec:num_diag} we explain how to perform numerical calculations for spin chains of modest size, but arbitrary $\gl(M|N)$ symmetry, using the walled Brauer algebra. Then we present numerical results on the spectrum.
Sec.~\ref{sec:rl} is of primary technical importance.
We discover a class of solutions to the $\gl(M|N)$ Bethe ansatz equations (BAE) which can be fully characterized in terms of solutions to the  $\gl(M-1|N-1)$ BAE. This provides an explicit embedding of the $\gl(M-1|N-1)$ spectrum into the $\gl(M|N)$ spectrum.
We give an algebraic explanation to this relationship.
Further, in sec.~\ref{sec:BAE} we identify the vacuum solution of $\gl(M|N)$ BAE and write it down explicitly in terms of the $\gl(M-N)$ vacuum solution. We then classify $\gl(M|N)$ excitations that lead to a $\gl(M-N)$ like spectrum and the simplest ones that do not. For the latter, we analyze numerically the form of the solutions to the BAE.
Finally, in sec.~\ref{sec:cl} we consider the continuum limit of spin chains with alternating inhomogeneities.
Excitations which can be characterized in terms of solutions to $\gl(M-N)$ BAE lead to a spectrum of massive particles with $\gl(M-N)$ Gross-Neveu mass ratios and $S$-matrix eigenvalues.
We argue that the continuum limit of the spin chain is the $\gl(M|N)$ Gross-Neveu model.
We end with analyzing some low lying excitations of the $\gl(2m|1)$ spin chain which are not of $\gl(2m-1)$ type and lead to new massive particles.

%%%%%%%%%%%%%%%%%%%%%%%%%%%%%%%%%%%%%%
%%%%%%%%%%%%%%%%%%%%%%%%%%%%%%%%%%%%%%

\section{Transfer matrices, Hamiltonians and their spectra}
\label{sec:formalism}

Let $V$ be the fundamental module of $\gl(M|N)$, $V^*$ denote its dual and $\rho:\gl(M|N)\mapsto \gl(V)$ together with $\bar{\rho}:\gl(M|N)\mapsto \gl(V^*)$ be the corresponding representations.
If $\{e_\alpha\}_{\alpha=1}^{M+N}$ is a graded basis of $V$ with grading $|\alpha|:=|e_\alpha|\in \mathbb{Z}/2\mathbb{Z}$ and $\{e^\alpha\}_{\alpha=1}^{M+N}$ is the dual basis $e^\alpha(e_\beta)=\delta_{\alpha\beta}$, then $E_{\alpha\beta}$ are the standard generators of $\gl(M|N)$ acting in $V$ as $E_{\alpha\beta}\cdot e_\gamma=\delta_{\beta\gamma}e_\alpha$ and in $V^*$ as $E_{\alpha\beta}\cdot e^\gamma=-(-1)^{(|\alpha|+|\beta|)|\alpha|}\delta_{\alpha\gamma}e^\beta$.
Let us label the $V$ factors of the spin chain $\mathcal{C}(L)=(V\otimes V^*)^{\otimes L}$ from left to right by a subscript $1,\dots,L$. Similarly, we label the $V^*$ factors  by a subscript $\bar{1},\dots,\bar{L}$.
For $\mathcal{E}\in \End V$ we denote by $\mathcal{E}_k\in\End\mathcal{C}(L)$  the endomorphisms acting as $\mathcal{E}$ on $V_k$ and trivially, up to grading signs, everywhere else in the chain. Similarly, for $\mathcal{E}\in \End V^*$, $\mathcal{E}_{\bar{k}}\in\End\mathcal{C}(L)$ will act as $\mathcal{E}$ on $V^*_{\bar{k}}$ and trivially, up to grading signs, everywhere else.   

The $\mathcal{C}(L)$--endomorphisms $P_{kl}=(-1)^{|\beta|}\rho_k(E_{\alpha\beta}) \rho_l(E_{\beta\alpha})$ provide a representation for the symmetric group acting on the $V$ factors of $\mathcal{C}(L)$. Similarly, $P_{\bar{k}\bar{l}}=(-1)^{|\beta|}\bar{\rho}_{\bar{k}}(E_{\alpha\beta}) \bar{\rho}_{\bar{l}}(E_{\beta\alpha})$
generate a representation for the symmetric group acting on the $V^*$ factors of $\mathcal{C}(L)$.
On the other hand, $Q_{k\bar{l}}= - (-1)^{|\beta|}\rho_k(E_{\alpha\beta}) \bar{\rho}_{\bar{l}}(E_{\beta\alpha})$ generate a representation of the Temperley-Lieb algebra $T_{2L}(n)$ with loop weight $n=M-N$.
Together, the $P$'s and $Q$'s generate the $\gl(M|N)$--centralizer algebra of $\mathcal{C}(L)$, that is the set of all endomorphisms of the spin chain that commute with the $\gl(M|N)$ action~\cite{Sergeev}.
This centralizer algebra is a representation of the walled Brauer algebra $B_{L,L}(n)$, which can be viewed as a subalgebra of the Brauer algebra $B_{2L}(n)$. We shall discuss in detail the algebra $B_{L,L}(n)$ and its representations in sec.~\ref{sec:wb_sub}.

In terms of walled Brauer algebra generators we introduce the $R$--matrices
\begin{align*}
R_{ij}(u)&= u+P_{ij}& R_{\bar{\imath}j}(u)&= u-Q_{\bar{\imath}j}\\
R_{i\bar{\jmath}}(u)&=u-Q_{i\bar{\jmath}}& R_{\bar{\imath}\bar{\jmath}}(u)&=u+P_{\bar{\imath}\bar{\jmath}}\ ,
\end{align*}
which satisfy the Yang-Baxter algebra
\begin{align}\label{eq:YBE}
 R_{ij}(u-v)R_{ik}(u)R_{jk}(v)&=R_{jk}(v)R_{ik}(u)R_{ij}(u-v) \\ \notag R_{ij}(u-v)R_{i\bar{k}}(u)R_{j\bar{k}}(v)& = R_{j\bar{k}}(v)R_{i\bar{k}}(u)R_{ij}(u-v)\\ \notag
 R_{i\bar{\jmath}}(u-v+n)R_{ik}(u)R_{\bar{\jmath}k}(v)&=R_{\bar{\jmath}k}(v)R_{ik}(u)R_{i\bar{\jmath}}(u-v+n)
\end{align}
and similar ``dual'' relations, that is the above relations with all barred indices replaced with unbarred ones and all unbarred indices with barred ones.
We define two one-parameter families of monodromies
\begin{align}\label{eq:mon_mat_a}
 T_a(u)&=R_{a\bar{L}}(u  +n/2)R_{aL}(u)\dots  R_{a\bar{1}}(u+n/2)R_{a1}(u)\\
  \bar{T}_{\bar{a}}(u)&= R_{\bar{a}\bar{L}}(u)R_{\bar{a}L}(u+n/2) \dots R_{\bar{a}\bar{1}}(u) R_{\bar{a}1}(u+n/2) \label{eq:mon_mat_ab}
\end{align}
acting on $V_a\otimes\mathcal{C}(L)$ and $V_{\bar{a}}\otimes \mathcal{C}(L)$, respectively,
and corresponding transfer matrices
\begin{equation}\label{eq:tmat}
 t(u)=\str_a T_a(u),\qquad \bar{t}(u)=\str_{\bar{a}}\bar{T}_{\bar{a}}(u)\ .
\end{equation}
%
%where $d$ is an exponentiated element of the Cartan subalgebra of $\un(M|N)$.
%The twist parameter $d$ and the inhomogeneities $u_j,\bar{u}_j$ have been introduced in order to properly deal later with degenerate solutions of Bethe ansatz equations.
Yang-Baxter relations~\eqref{eq:YBE} imply a Yangian structure given by the two relations
\begin{align}\label{eq:comm1}
 R_{ab}(u-v)T_a(u)T_b(v)&=T_b(v)T_a(u)R_{ab}(u-v)\\
R_{a\bar{b}}(u-v+n/2)T_a(u)\bar{T}_{\bar{b}}(v)&=\bar{T}_{\bar{b}}(v)T_a(u)R_{a\bar{b}}(u-v+n/2)\label{eq:comm2}
\end{align}
and their duals.
The following commutation relations immediately follow
\begin{equation*}
 [t(u),t(v)]=[t(u),\bar{t}(v)]=[\bar{t}(u),\bar{t}(v)]=0\ .
\end{equation*}
%

%Due to duality relations of the type $Q_{a\bar{b}}=P_{ab}^{\dagger_b}$, the following relation between the two families of transfer matrices
%
%\begin{equation*}
 %\bar{t}(u)=t(-u^*-\lambda-n/2)^\dagger
%\end{equation*}
%
%exists, where $\dagger$ is the conjugate supertranspose and we have assumed that all inhomogeneities $u_j,\bar{u}_j$ and $\lambda$ are purely imaginary.
%Thus, solving the diagonalization  problem for $t(u)$ also yields the eigenvalue of $\bar{t}(u)$ on a given eigenvector of $t(u)$.

The nested algebraic Bethe ansatz for the most general $\gl(M|N)$ spin chain was considered in \cite{Belliard:2008di}. The Bethe ansatz equations and the spectrum of $t(u)$ formally depend on the nesting order, that is an ordering of a basis of $V$ and the induced ordering of the dual basis of $V^*$.
If  the basis $\{e_\alpha\}_{\alpha=1}^{M+N}$ of $V$  diagonalizes the Cartan subalgebra, then, without loss of generality, we label the basis vectors so that the total ordering reads
\begin{equation}\label{eq:tot_ordering_bas_vecs}
 e_1>e_2>\dots>e_{M+N}\ ,
\end{equation}
where, however, we keep unspecified the grading of the basis vectors.
Let $\wt(e_\alpha)=\epsilon_\alpha$ denote the weights of basis vectors called \emph{fundamental weights}. The ordering~\eqref{eq:tot_ordering_bas_vecs} induces a weight space ordering $\epsilon_1>\dots>\epsilon_{M+N}$
which fixes the simple root system $\Delta_0=\{\alpha_i:=\epsilon_i-\epsilon_{i+1}\}_{i=1}^{M+N-1}$.
The choice of grading, which we denote by $\Sigma=\{\sigma_\alpha = |e_\alpha|=|\alpha|\}_{\alpha=1}^{M+N}$, is equivalent to the choice of a Cartan matrix, or a Dynkin diagram.
Changing the grading can be equivalently seen as changing the total ordering~\eqref{eq:tot_ordering_bas_vecs}. As a result, the simple root system, the positive root system and the Borel subalgebra changes with $\Sigma$.
So, keep in mind that the notion of highest weight always depends on the grading choice and changes when $\Sigma$ changes.

Define the operator matrix elements of the monodromies~(\ref{eq:mon_mat_a}, \ref{eq:mon_mat_ab}) as
$T = E_{ij}\otimes T_{ij}$, where $T_{ij}\in \End\mathcal{C}(L)$ and $T=T_{a},T_{\bar{a}}$. 
Choosing the reference state $\Omega$ to be the highest weight state of $\mathcal{C}(L)$, the eigenvalues of $t(u)$ can be written in terms of polynomials $(T_a)_{ii}(u)\Omega=\Lambda_i(u)\Omega$
\begin{equation*}
 \Lambda_i(u)=\begin{cases}
(u+(-1)^{|1|})^L(u+n/2)^L, & i=1\\
 u^L(u+n/2)^L, & 2\leq i\leq r\\
u^L(u+n/2-(-1)^{|M+N|})^L,& i=M+N
\end{cases}
\end{equation*}
and simple root $Q$-polynomials
\begin{equation*}
 Q_k(u)=\prod_{j=1}^{\nu^{(k)}}(u-u^{(k)}_j)
\end{equation*}
as follows
\begin{equation}\label{eq:spec_trmat}
 \Lambda(u)=\sum_{k=1}^{M+N} (-1)^{|k|}\Lambda_k(u) \frac{Q_{k-1}(u+(-1)^{|k|})Q_k(u-(-1)^{|k|})}{Q_{k-1}(u)Q_{k}(u)}
\end{equation}
where the $u^{(k)}_j$, $k=1,\dots,r=M+N-1$ are the Bethe roots appearing at the $k$--th step of the nesting. There are solutions of the following system of nested Bethe ansatz equations
\begin{equation}\label{eq:BAE_Qform}
\frac{\Lambda_{k}(u^{(k)}_j)}{\Lambda_{k+1}(u^{(k)}_j)}=-(-1)^{|k|+|k+1|}\frac{Q_{k-1}(u^{(k)}_j)Q_k(u^{(k)}_j+(-1)^{|k+1|})Q_{k+1}(u^{(k)}_j-(-1)^{|k+1|})}{Q_{k-1}(u^{(k)}_j+(-1)^{|k|})Q_k(u^{(k)}_j-(-1)^{|k|})Q_{k+1}(u^{(k)}_j)}\ ,
\end{equation}
ensuring the analyticity of eigenvalues~\eqref{eq:spec_trmat}, where $k=1,\dots,r$ and $j=1,\dots,\nu^{(k)}$ and it has to be understood that $Q_0(u)=Q_{M+N}(u)=1$. We stress that the BAE~\eqref{eq:BAE_Qform} are equivalent to the analyticity requirement of the transfer matrix~\eqref{eq:spec_trmat} if and only if none of the Bethe roots of the same type coincide, which is an essential requirement in the algebraic Bethe ansatz construction.

The BAE take a more familiar look~\cite{Ogievetsky:1986hu}
\begin{equation}\label{eq:BAE_short}
 \prod_{k=1}^{L} e_{\langle \Lambda_{k},\alpha \rangle }(x_j^{(\alpha)}-y_k)e_{\langle \Lambda_{\bar{k}},\alpha \rangle }(x_j^{(\alpha)}-y_{\bar{k}})=-(-1)^{|\alpha|}\prod_{\beta\in\Delta_0}\prod_{i=1}^{\nu^{\alpha}}e_{\langle\alpha,\beta\rangle}(x_j^{(\alpha)}-x_i^{(\beta)})\ ,
\end{equation}
when written in terms of new variables
\begin{equation}\label{eq:shifts}
 iu^{(k)}_{j}= x^{(k)}_j - \frac{i}{2}\sum_{l=1}^k(-1)^{|l|}\ ,
\end{equation}
the Takahashi functions
\begin{equation*}
 e_t(x)=\frac{x+it/2}{x-it/2}\, ,
\end{equation*}
the highest weights $\Lambda_k=\epsilon_1$ and $\Lambda_{\bar{k}}=-\epsilon_{M+N}$ of modules $V_k$ and $V^*_{\bar{k}}$ in $\mathcal{C}(L)$, 
and the weight space scalar product $\langle \epsilon_i, \epsilon_j\rangle = \delta_{ij}(-1)^{|i|}$.
The degree of a root $\alpha_k=\epsilon_k-\epsilon_{k+1}$ is defined as $|\alpha_k|=|k|+|k+1|$.
We shall be mostly considering the homogeneous case~(\ref{eq:mon_mat_a}, \ref{eq:mon_mat_ab}) corresponding to 
$y_k = y_{\bar{k}}= 0$, although the inhomogeneous deformation $y_k,y_{\bar{k}}\neq 0$ shall also be required.
There are $j=1,\dots,\nu^{\alpha}$ equations for every $\alpha\in \Delta_0$.

The weight of the reference state is $\wt(\Omega)=\sum_{k=1}^L\Lambda_k+\sum_{\bar{k}=1}^L\Lambda_{\bar{k}}=L(\epsilon_1-\epsilon_{M+N})$. Bethe vectors $\omega$ described by the Bethe roots~\eqref{eq:BAE_short} are highest weight vectors of weight
\begin{equation}
 \wt(\omega) = \wt(\Omega) - \sum_{k=1}^r\alpha_k \nu^{(k)}\ .
\label{eq:weight_BV1}
\end{equation}

We define the dynamics of the spin chain by the momentum and Hamiltonian operators
\begin{equation}\label{eq:first_charges}
 H = \frac{d}{du}\bigg\vert_{u=0}\log \frac{t(u)\bar{t}(u)}{\Lambda_1(u)\bar{\Lambda}_{M+N}(u)}\ ,\qquad \exp{i P} = (-1)^{|1|+|M+N|}\frac{t(0)\bar{t}(0)}{\Lambda_1(0)\bar{\Lambda}_{M+N}(0)}\ .
\end{equation}
In order to have explicit expression for the spectrum of these operators, the eigenvalues~\eqref{eq:spec_trmat} of $t(u)$ are not enough. One also needs to evaluate the eigenvalue of $\bar{t}(u)$ on a given Bethe eigenvector of $t(u)$.
A \emph{fundamental difference} w.r.t. $\gl(N)$ spin chains is that one cannot solve this problem by fusion. This is because tensor products of the fundamental representation $V$ of $\sgl(M|N)$ will never generate the dual representation $V^*$ as a direct summand, nor even as a subquotient.
To develop a clear idea about how this should be done, let us recall that a set of Bethe vectors for $t(u)$ can be constructed in the framework of the algebraic Bethe ansatz (ABA) by using only the commutation relations~\eqref{eq:comm1}. Then, the eigenvalue of $\bar{t}(u)$ on such a Bethe vector can be calculated, in principle, by using the second type of commutation relations~\eqref{eq:comm2}.
We shall not pursue this rather tedious route. Instead, we guess the eigenvalue of $\bar{t}(u)$ on a given Bethe vector of $t(u)$ as follows.

First, notice that a different set of Bethe vectors can be obtained by performing the ABA for $\bar{t}(u)$, that is by using the commutation relations dual to eq.~\eqref{eq:comm1}.
We perform the nesting by ordering the dual basis vectors $\{e^\alpha\}_{\alpha=1}^{M+N}$ of the auxiliary space $V^*$ according to their weights $\wt(e^\alpha)=-\epsilon_\alpha$
\begin{equation}
 e^{M+N}>\dots >e^2>e^1\ .
\label{eq:dual_ord}
\end{equation}
We denote the Bethe roots appearing at step $k$ of the nested ABA by $\bar{u}^{(M+N-k)}_j$, because the simple root which must be associated to them is $\wt(e^{M+N-k+1})-\wt(e^{M+N-k})=\alpha_{M+N-k}$.
Then, the eigenvalues of $\bar{t}(u)$ can be written in term of polynomials 
$(T_{\bar{a}})_{ii}(u)\Omega=\bar{\Lambda}_i(u)\Omega$
\begin{equation*}
 \bar{\Lambda}_i(u)=\begin{cases}
(u-(-1)^{|1|}+n/2)^L u^L, & i=1\\
 (u+n/2)^Lu^L, & 2\leq i\leq r\\
(u+n/2)^L(u+(-1)^{|M+N|})^L,& i=M+N
\end{cases}
\end{equation*}
and simple root polynomials $\bar{Q}_k(u)=\prod_{j=1}^{\bar{\nu}^{(k)}}(u-\bar{u}^{(k)}_j)$
in complete analogy with~\eqref{eq:spec_trmat}
\begin{equation}
\bar{\Lambda}(u)=\sum_{k=1}^{M+N}(-1)^{|k|}\bar{\Lambda}_k(u)\frac{\bar{Q}_{k-1}(u-(-1)^{|k|})\bar{Q}_k(u+(-1)^{|k|})}{\bar{Q}_{k-1}(u)\bar{Q}_k(u)}\ .
\label{eq:eg_trmat_bar}
\end{equation}
Analyticity conditions for $\bar{\Lambda}(u)$ take the same form as eqs.~\eqref{eq:BAE_short} in terms of variables
\begin{equation}
 \bar{u}^{(k)}_j = \bar{x}^{(k)}_j-\frac{i}{2}\sum_{l=k+1}^{M+N}(-1)^{|l|}
\label{eq:bshifts}
\end{equation}
and parameters $\bar{\nu}^{(k)}$.
A Bethe vector constructed in this way, which we denote by $\bar{\omega}$, has weight 
\begin{equation*}
\wt(\bar{\omega})=\wt(\Omega)-\sum_{k=1}^r \alpha_k \bar{\nu^{(k)}}\ .
%\label{eq:}
\end{equation*}

At this point, the eigenvalues of $t(u)$ w.r.t. the second set of Bethe vectors $\bar{\omega}$ is not known. If we can match Bethe vectors $\omega$ with Bethe vectors $\bar{\omega}$ then the problem is solved.
Due to the subtle completeness issue of ABA for super spin chains, it is not clear at all if the matching can actually be performed.
We assume it can and we do it as follows: a Bethe vector $\omega$ of $t(u)$ is identified with a Bethe vector $\bar{\omega}$ of $\bar{t}(u)$ if
\begin{align}\label{eq:indent_BV}
&&\omega =& \bar{\omega} & {}&\Leftrightarrow&
\begin{cases}
   \nu^{(k)}=\bar{\nu}^{(k)}\\ \{x^{(k)}_j\}_{j=1}^{\nu^{(k)}}=\{\bar{x}^{(k)}_j\}_{j=1}^{\bar{\nu}^{(k)}} &
 \end{cases} &  k=1,\dots r \ .
\end{align}
The first condition in the braces means $\wt(\omega)=\wt(\bar{\omega})$.
Eqs.~(\ref{eq:spec_trmat}, \ref{eq:eg_trmat_bar}, \ref{eq:shifts}, \ref{eq:bshifts}) allow us to compute both the eigenvalues of $t(u)$ and $\bar{t}(u)$ on a Bethe vector $\omega=\bar{\omega}$ in~\eqref{eq:indent_BV}.

After this long detour, we come back to the spectrum of the operators~\eqref{eq:first_charges}, which can now be explicitly computed
\begin{align}\label{eq:spec_ham_mom}
 E &= -\sum_{j=1}^{\nu^{(1)}}\frac{(-1)^{|1|}}{x^{(1)}_j {}^2+1/4}-\sum_{j=1}^{\nu^{(r)}}\frac{(-1)^{|M+N|}}{x^{(r)}_j{}^2+1/4}\\
P &\equiv  \sum_{i=1}^{\nu^{(1)}}(-1)^{|1|}\theta_1(x^{(1)}_j)+ \sum_{i=1}^{\nu^{(r)}}(-1)^{|M+N|}\theta_1(x^{(r)}_j)\mod 2\pi \ ,
\end{align}
where $\theta_t(u)=i \log e_t(u)+\pi=2\tan^{-1}\tfrac{2u}{t}$ with some fixed branch.
First thing to be noticed is the explicit dependence of the definitions~\eqref{eq:first_charges} on the grading.
Therefore, it looks like the type of the chain -- ferromagnetic or antiferromagnetic -- also depends on it, which is also suggested by the grading signs in \eqref{eq:spec_ham_mom}.
However, the charges $H$ and $P$ can be written explicitly as a representation of an element of the periodic walled Brauer algebra
\begin{align}\label{eq:ham}
 H &= \sum_{i=1}^L -(-1)^{|1|} -(-1)^{|M+N|}+ P_{ii+1}+ P_{\overline{i-1i}} -\frac{2}{n}\left(\{ Q_{\bar{i}i+1},Q_{i\bar{i}}\} + \{Q_{\overline{i-1}i},Q_{i\bar{i}}\}\right)
\end{align}
where all (un)barred indices are defined modulo $L$ and the affine generators are expressed in terms of non-periodic walled Brauer algebra elements $P_{L1}:=P_{1L}=P_{12}\dots P_{L-1L}\dots P_{12}$,  $P_{\overline{L1}}:=P_{\overline{1L}}=P_{\overline{12}}\dots P_{\overline{L-1L}}\dots P_{\overline{12}}$ and $Q_{\bar{L}1}:=Q_{1\bar{L}} = P_{1L}Q_{L\bar{L}}P_{1L}$.
The following general formulas have been used
\begin{align*}
 \frac{d\log t(0)}{du}&=\sum_{i=1}^L R^{-1}_{i\bar{i}}\left(\frac{n}{2}\right)\dot{R}_{i\bar{i}}\left(\frac{n}{2}\right)+\sum_{i=1}^{L} R^{-1}_{i\bar{\imath}}\left(\frac{n}{2}\right)\dot{\check{R}}_{ii+1}(0)R_{i\bar{\imath}}\left(\frac{n}{2}\right)\\
  \frac{d\log \bar{t}(0)}{du}&=\sum_{i=1}^L\dot{R}_{i\bar{i}}\left(\frac{n}{2}\right)R^{-1}_{i\bar{i}}\left(\frac{n}{2}\right)+\sum_{i=1}^{L}R_{i\bar{\imath}}\left(\frac{n}{2}\right)\dot{\check{R}}_{\overline{i\imath-1}}(0)R^{-1}_{i\bar{i}}\left(\frac{n}{2}\right)\ .
\end{align*}
The latter can be derived using only the cyclicity of the supertrace, which holds for a graded tensor product when even endomorphisms are considered, relations of the form $P_{ij}R_{jk}(u)= R_{ik}(u)P_{ij}$, $P_{ij}R_{j\bar{k}}(u)= R_{i\bar{k}}(u)P_{ij}$, $\str_a P_{aj} = \id$ and their duals.
We see from eq.~\eqref{eq:ham} that the grading enters in the definition of the Hamiltonian only as a shift, therefore having nothing to do with ferro or antiferromagnetism.
The equivalence of the solutions of BAE in different gradings, and therefore of $\spec H$, is at present somewhat understood in terms of particle hole transformations~\cite{Tsuboi:1998ne}, although the relationship between Bethe vectors in different gradings not at all.

A relation between the spectrum of $H$ and $-H$ can be constructed, but one has to consider different chains. Let $H_{M|N}$ denote the integrable Hamiltonian of the $\gl(M|N)$ spin chain. Then on has the following relation
\begin{equation}\label{eq:aut}
 H_{M|N} = - H_{N|M} \ .
\end{equation}
To prove it one uses the algebra homomorphism between the walled Brauer algebras $B_{L,L}(n)$ and $B_{L,L}(-n)$ provided by
$P\mapsto -P$ and $Q\mapsto -Q$. This automorphism is realized in the spin chain representation by the shift of the grading function $|i|\mapsto |i|+1$, which  maps $\gl(M|N)\mapsto \gl(N|M)$.
As we shall see in the next section, the sign in front of our Hamiltonian~\eqref{eq:ham} ensures that we are dealing with antiferromagnetic spin chains for $n=M-N>0$.
Eq.~\eqref{eq:aut} allows to fold back the $\gl(N|M)$ spin chains with $N<M$ to $\gl(M|N)$ spin chains with $n>0$ by changing the sign of the Hamiltonian~\eqref{eq:ham}, but now they will be ferromagnetic.
As the  $\gl(N|N)$ chain in eq.~\eqref{eq:ham} is poorly defined, we shall discard it from the main discussion and come back to it only in the conclusions.
From now on we restrict to $\gl(M|N)$ antiferromagnetic spin chains~\eqref{eq:ham} for which $n=M-N>0$.

%%%%%%%%%%%%%%%%%%%%%%%%%%%%%%%%%%%%%%%%%%%%%%%%%%%%%%
%%%%%%%%%%%%%%%%%%%%%%%%%%%%%%%%%%%%%%%%%%%%%%%%%%%%%%
%%%%%%%%%%%%%%%%%%%%%%%%%%%%%%%%%%%%%%%%%%%%%%%%%%%%%%
\section{Numerical diagonalization of the Hamiltonian}
\label{sec:num_diag}
%%%%%%%%%%%%%%%%%%%%%%%%%%%%%%%%%%%%%%%%%%%%%%%%%%%%%%
%%%%%%%%%%%%%%%%%%%%%%%%%%%%%%%%%%%%%%%%%%%%%%%%%%%%%%
%%%%%%%%%%%%%%%%%%%%%%%%%%%%%%%%%%%%%%%%%%%%%%%%%%%%%%

Diagonalizing the matrix~\eqref{eq:ham} is not the smartest thing to do if one is solely interested in its spectrum, especially if one intends to consider on the same footing all the $\gl(M|N)$--spin chains. This is because many eigenvalues have multiplicities corresponding to the dimension of irreducible $\gl(M|N)$ representations appearing (generically as quotients of submodules) in the spin chain. These multiplicities quickly grow with the rank and the computing power per eigenvalue increases respectively.
An approach which allows to select the eigenvalues corresponding to a given $\gl(M|N)$--irreducible representation and eliminate the corresponding degeneracy would be considerably more efficient.
To implement this approach one interprets the Hamiltonian $H$ in eq.~\eqref{eq:ham} as an element $\mathsf{H}$ of some algebra, namely the walled Brauer algebra $B_{L,L}(n=M-N)$, in a particular representation provided by the $\gl(M|N)$--centralizer of the $\mathcal{C}(L)$ spin chain.
As we shall explain shortly, the algebra $B_{L,L}(n)$ can be abstractly defined independently of the $\gl(n+N|N)$ spin chains and, in particular, it does not depend on $N$. Centralizers of
$\gl(n+N|N)$ spin chains provide $N$ dependent representations for $B_{L,L}(n)$.
The important thing is that the representation theory of $B_{L,L}(n)$ is understood well enough, so that one can find the spectrum of the algebraic Hamiltonian $\mathsf{H}$ by working in other representations where its spectrum is much less
%~\footnote{We shall work with the standard modules of $B_{L,L}(n)$, which are always reducible for $L$ big enough. Therefore, there will still be residual degeneracies. These, however, will not depend on the rank $M+N-1$, but rather on $n=M-N$.}
degenerate compared to that in the $\gl(n+N|N)$ spin chain, namely it does not depend on $N$.

%%%%%%%%%%%%%%%%%%%%%%%%%%%%%%%%%%%%%%%%%%%%%
%%%%%%%%%%%%%%%%%%%%%%%%%%%%%%%%%%%%%%%%%%%%%
\subsection{Walled Brauer algebra and its standard modules}
\label{sec:wb_sub}
%%%%%%%%%%%%%%%%%%%%%%%%%%%%%%%%%%%%%%%%%%%%%
%%%%%%%%%%%%%%%%%%%%%%%%%%%%%%%%%%%%%%%%%%%%%

\begin{figure}[t]%
\psfrag{1}{$1$}
\psfrag{l}{$L$}
\psfrag{d}{$\cdots$}
\psfrag{i}{$i$}
\psfrag{i}{$i$}
\psfrag{P}{$\mathsf{P}_i$}
\psfrag{I}{$\mathsf{I}$}
\psfrag{E}{$\mathsf{E}_i$}
\psfrag{PPP}{$\mathsf{P}_{i-1}\mathsf{P}_{i}\mathsf{P}_{i+1}$}
\centerline{\includegraphics[scale=0.8]{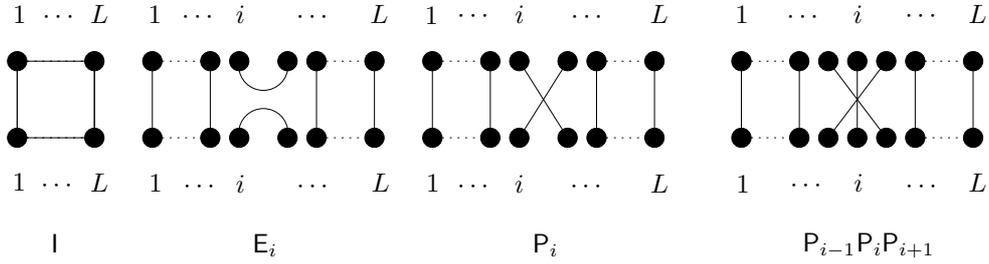}}\caption{The identity $\mathsf{I}$ and the generators $\mathsf{E}_i,\mathsf{P}_i$ of the Brauer algebra of
dimension
$(2L-1)!!$ are represented on the left; the walled Brauer algebra generator
$\mathsf{P}_{i-1} \mathsf{P}_{}\mathsf{P}_{i+1}$
is represented on the right.}%
\label{fig:gen_br_alg}%
\end{figure}

The walled Brauer algebra $B_{L,L}(n)$ can be conveniently viewed as a subalgebra of the Brauer algebra $B_{2L}(n)$.
The defining relations of $B_{2L}(n)$ can be found in \cite{ram}.
The words of $B_{2L}(n)$ admit a representation as graphs
on $4L$ labeled vertices with $2L$ edges connecting the vertices
pairwise in all $(4L-1)!!$ possible ways (crossings are allowed).
The identity $\mathsf{I}$ of the Brauer algebra and the generators $\mathsf{E}_i,\mathsf{P}_i$
are represented by the graphs on the left in
fig.~\ref{fig:gen_br_alg}.
%
%\begin{figure}[t]%
%\centerline{\includegraphics[scale=0.8]{fig/gen_diag.eps}
%\hspace{2cm}
%\includegraphics[scale=0.8]{fig/ad_gen.eps}}%
%\caption{The identity $\mathsf{I}$ and the generators $\mathsf{E}_i,\mathsf{P}_i$ of the Brauer algebra of
%dimension
%$(2L-1)!!$ are represented on the left; the walled Brauer algebra generator
%$\mathsf{P}_i \mathsf{P}_{i+1}\mathsf{P}_i$
%is represented on the right.}%
%\label{fig:gen_br_alg}%
%\end{figure}

%

In order to multiply the diagrams one arranges the first $2L$
vertices horizontally with the remaining $2L$ vertices on top of
the first ones. The product of a diagram $d_1$ with a diagram
$d_2$ is the diagram $d_1 d_2$ obtained by i) placing the diagram
$d_1$ on top of the diagram $d_2$, ii) identifying the top of the
diagram $d_2$ with the bottom of the diagram $d_1$ and iii)
replacing every loop generated in this process by $n$.
%The periodic
%Brauer algebra is an extension of the Brauer algebra by two
%generators $E_{2L}$ and $P_{2L}$ which satisfy the same defining
%relation as the generators of the Brauer algebra if the index
%$i\equiv i+2L$ is regarded as periodic. The words of the periodic
%Brauer algebra are diagrams with the top and the bottom being
%circles wrapped around a cylinder and carrying $2L$ vertices each,
%such that the latter are pairwise connected in all possible ways
%by $2L$ edges  living on the surface of the cylinder. The periodic
%Brauer algebra has infinite dimension.
The walled Brauer algebra $B_{L,L}(n)$ is the subalgebra of $B_{2L}(n)$ generated by the elements $\mathsf{Q}_{i\bar{\imath}}:=\mathsf{E}_{2i-1}$, $\mathsf{Q}_{\bar{\imath}i+1}:=\mathsf{E}_{2i}$, $\mathsf{P}_{i,i+1}:= \mathsf{P}_{2i-1} \mathsf{P}_{2i}\mathsf{P}_{2i-1}$, $\mathsf{P}_{\overline{\imath\imath+1}}:= \mathsf{P}_{2i} \mathsf{P}_{2i+1}\mathsf{P}_{2i}$. The generators
$\mathsf{P}_i\mathsf{P}_{i+1}\mathsf{P}_i$ are represented on the right in
fig.~\ref{fig:gen_br_alg}.
For every diagram spanning $B_{2L}(n)$ with vertices labeled as in fig.~\ref{fig:gen_br_alg}, imagine moving all odd vertices to the left of a wall, all the even ones to the right, while keeping the connectivity unchanged. Then $B_{L,L}(n)$ is spanned by the set of $B_{2L}(n)$ diagrams, such that only strictly horizontal edge cross the wall.
In this representation of $B_{L,L}(n)$ we label the  $L$ up and $L$ down vertices on the left of the wall by the set $1,\dots,L$ from left to right and similarly those on the right by the set $\bar{1},\dots,\bar{L}$.
The $\mathsf{P}_{ii+1}$ generators act on the left of the wall, the $\mathsf{P}_{\overline{\imath\imath+1}}$ act on the right, while the  generators  $\mathsf{Q}_{i\bar{\imath}}$ and $\mathsf{Q}_{\bar{\imath}i+1}$ act across the wall.

\begin{figure}%
\psfrag{x}{$\times$}
\psfrag{o}{$\otimes$}
\psfrag{v}{$\mathbf{v}$}
\psfrag{u}{$((2,1), (1,2))\cdot \mathbf{v}$}
\psfrag{1}{$1$}
\psfrag{2}{$2$}
\psfrag{f}{$\phantom{fffff}$}
\centerline{\includegraphics[scale=0.7]{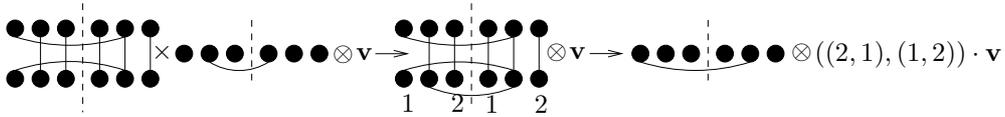}}
\caption{Example of diagrammatic multiplication of a basis element of $\Delta_{3,3}(\lambda,\mu)$, where $\lambda,\mu\vdash 2$ and $\mathbf{v}$ is an element of the module $S(\lambda)\times S(\mu)$ of $\Sym(2)\times\Sym(2)$.}
\label{fig:mult_diag}
\end{figure}

Next we give a brief description of a set of modules of $B_{L,L}(n)$ over which we actually numerically diagonalize the algebraic Hamiltonian $\mathsf{H}$.
These modules shall be related in the following to $\gl(n+N|N)$ \emph{traceless tensors} of fixed co-- and contravariant shapes.
For $\lambda$ and $\mu$ partitions of the non-negative integers $|\lambda|,|\mu|=f\leq L$, we denote by $\Delta_{L,L}(\lambda,\mu)$ the \emph{standard modules} of $B_{L,L}(n)$.
There are constructed in the following way.
Let $\Sym(f)$ denote the symmetric group on $f$ objects and $S(\lambda)$, $S(\mu)$ denote its simple modules labeled by the corresponding partitions.
Then, $\Delta_{L,L}(\lambda,\mu)$ has as basis the tensor products between the set of some diagrams on $2L$ points, with $L$ of them on the either side of the wall, and some basis of $S(\lambda)$ and $S(\mu)$.
The diagrams are such that every point is either free or belongs to a horizontal edge crossing the wall. The number of edges is fixed to $L-f$.
We give a rough idea about how the action of the walled Brauer algebra can be constructed by diagrammatic multiplication in in fig.~\ref{fig:mult_diag}: i) in a first step, before the diagrammatic multiplication, assign labels to the free points of the diagrams on each side of the wall; ii) in a second step, after the diagrammatic multiplication, apply a surjective homomorphism from labeled diagrams to the tensor product of unlabeled diagrams with $S(\lambda)$ and $S(\mu)$, that is to $\Delta_{L,L}(\lambda,\mu)$. All labeled diagrams with more then $L-f$ horizontal edges that can appear as a result of the diagrammatic multiplication belong to the kernel of this homomorphism.
In particular, $\Delta_{f,f}(\lambda,\mu)\simeq S(\lambda)\times S(\mu)$.
The detailed definitions can be found in \cite{martin}.

We implement the action of $B_{L,L}(n)$ in $\Delta_{L,L}(\lambda,\mu)$ on a computer and investigate the spectrum of the algebraic Hamiltonian $\mathsf{H}$. Before presenting and discussing the numerical results of sec.~\ref{sec:spec} one should explain how to extract from these data the spectrum of the original spin chain Hamiltonian $H$.

%%%%%%%%%%%%%%%%%%%%%%%%%%%%%%%%%%%%%%%%%%%%%%%%%%%%%%%%%%
%%%%%%%%%%%%%%%%%%%%%%%%%%%%%%%%%%%%%%%%%%%%%%%%%%%%%%%%%%
\subsection{Traceless tensors}\label{sec:tens}
%%%%%%%%%%%%%%%%%%%%%%%%%%%%%%%%%%%%%%%%%%%%%%%%%%%%%%%%%%
%%%%%%%%%%%%%%%%%%%%%%%%%%%%%%%%%%%%%%%%%%%%%%%%%%%%%%%%%%

Consider the $\gl(M|N)$ tensor $(V\otimes V^*)^{\otimes f}$.
There is an obvious action of $\Sym(f)\times\Sym(f)$ on the $V$ and $V^*$ factors.
Let $\lambda,\mu$ be partitions of $f$, which we symbolically write as $\lambda,\mu\vdash f$.
One can apply a Young symmetrizer of shape $\lambda$ to the $V$ factors and a Young symmetrizer of shape $\mu$ to the $V^*$ factors to get a tensor of shape $t(\lambda,\mu)$ \cite{Weyl}.
We say that $t(\lambda,\mu)$ has rank $(f,f)$, covariant shape $\lambda$ and contravariant shape $\mu$ or, shortly, shape $(\lambda,\mu)$.
The number of Young symmetrizers of shape $\lambda$ is equal to the 
number of standard Young tableau of shape $\lambda$, which is also equal to $\dim S(\lambda)$.
Therefore, the number of tensors of shape $(\lambda,\mu)$ is $\dim S(\lambda)\times \dim S(\mu)$.
Every tensor $t(\lambda,\mu)$ is a $\gl(M|N)$ module, which appears as a direct summand in $(V\otimes V^*)^{\otimes f}$.
The symmetric group $\Sym(f)\times\Sym(f)$ acts in the space of tensors of the same shape $(\lambda,\mu)$, transforming them into each other. The latter subspace is isomorphic to a direct sum of $S(\lambda)\otimes S(\mu)$ modules of $\Sym(f)\times\Sym(f)$. 
These statements can be compactly written as follows
\begin{equation*}
(V\otimes V^*)^{\otimes f}\mathop{\simeq}_{\Sym(f)\times \Sym(f)}\bigoplus_{\lambda,\mu \vdash f}\dim t(\lambda,\mu)S(\lambda)\otimes S(\mu)\ .
%\label{eq:}
\end{equation*}
%

% We believe the allowable shapes are
%
%\begin{equation}\label{eq:admiss_shape}
%t(\lambda,\mu)\neq 0\qquad \Leftrightarrow \qquad r(\lambda)+r(\mu)\leq M\, ,\; c(\lambda)+c(\mu)\leq N\ ,
%\end{equation}
%
%which reduce to well known $(M,N)$--hook shapes for purely co-- or contravariant tensors and to staircases for $\gl(n)$ groups \cite{}.

Consider now the subspace $t_0(\lambda,\mu)\subset t(\lambda,\mu)$ of traceless tensors. Notice that this is a $\gl(M|N)$
submodule, which will  \emph{not} necessarily be \emph{a direct summand} of $t(\lambda,\mu)$.
More generally, one can consider the $\gl(M|N)$ submodules $t_{n-1}(\lambda,\mu)\subset t(\lambda,\mu)$ composed of tensors whose all contractions of $n$ covariant indices with $n$ contravariant indices vanish.
These provide a filtration of the tensor $t(\lambda,\mu)$
\begin{equation}
 t_0(\lambda,\mu)\subset t_1(\lambda,\mu)\subset \cdots \subset t_f(\lambda,\mu)=t(\lambda,\mu)\ .
\label{eq:filtr}
\end{equation}
It is very important to observe that the subquotients $t_{n}(\lambda,\mu)/t_{n-1}(\lambda,\mu)$ of this filtration are isomorphic to traceless tensors of lower rank $(f-n,f-n)$.
For instance, taking a trace of $t_1(\lambda,\mu)$ one gets a traceless tensor of rank $f-1$ because, by definition, all double contractions of $t_1(\lambda,\mu)$ must vanish. Now, the preimage of single traces of $t_1(\lambda,\mu)$ modulo the kernel $t_0(\lambda,\mu)$ of the single trace homomorphisms is precisely to $t_1(\lambda,\mu)/t_0(\lambda,\mu)$. Therefore, $t_1(\lambda,\mu)/t_0(\lambda,\mu)$ is isomorphic to traceless tensors of rank $(f-1,f-1)$.

We see that traceless tensors $t_0(\lambda,\mu)$ have a fundamental role --- all direct summands of the spin chain $(V\otimes V^*)^{\otimes L}$ can be built out of them.
As a consequence, the full spectrum of the spin chain Hamiltonian can be reconstructed from the spectra in the subspaces of traceless tensors of shape $\lambda,\mu\vdash f$, $f=0,1,\dots,L$.
In fact, not all of these shapes are possible. We shall determine the class of shapes for which the traceless tensors do not vanish later.

The traceless tensors are not necessarily irreducible. One way to build submodules of a traceless tensor $t_0(\lambda,\mu)$ is by embedding into it quotients of traceless tensors of lower rank as follows. Let $t_0(\lambda',\mu')$ be a traceless tensor, $\lambda',\mu'\vdash f-k$, $\lambda'\subset \lambda$, $\mu'\subset \mu$ and $\mathsf{e}_{\lambda},\mathsf{e}_\mu$ denote some Young symmetrizers of shape $\lambda$, $\mu$. The tensor  
\begin{equation*}
\mathsf{e}_{\lambda}\mathsf{e}_\mu t_0(\lambda',\mu')\otimes \left((V\otimes V^*)^{\otimes k}\right)^{\gl(M|N)} \subset t(\lambda,\mu)
\end{equation*}
might have an intersection with a non-trivial submodule of $t_0(\lambda,\mu)$. The latter will generally be isomorphic to only a quotient of $t_0(\lambda',\mu')$, because the Young symmetrizers $\mathsf{e}_{\lambda},\mathsf{e}_\mu$ are projectors.
An illustrative example is the indecomposable $\gl(N|N)$ tensor $t(1,1)= V\otimes V^*$. The traceless subspace $t_0(1,1)$, isomorphic to the adjoint representation, is spanned by elements of the form $G^i_j e_i\otimes e^j$ subject to the constraint $\str G = G^i_i (-1)^{|i|}=0$, where $\{e_i\}_{i=1}^{2N}$ is a basis of $V$ and $\{e^i\}_{i=1}^{2N}$ is the dual basis. The quotient $t_1(1,1)/t_0(1,1)$ is one dimensional. A coset representative for this quotient is, for instance, $(-1)^{|i|}e_i\otimes e^i$.
The traceless tensor $t_0(1,1)$ is also indecomposable, but reducible.
It has a unique proper non-trivial submodule  spanned by the $\gl(N|N)$ traceless invariant $e_i\otimes e^i$.

Represent now the covariant part of every tensor $t_0(\lambda,\mu)$ of fixed shape $(\lambda,\mu)$ and ranks $(f,f)$ by $f$ dots on the left of an imaginary wall and the contravariant part by $f$ dots on the right of that wall.
Then the walled Brauer algebra generators $\mathsf{P}_{ii+1}$ will act on the left of the wall as in $S(\lambda)$ and the $\mathsf{P}_{\overline{\imath\imath+1}}$ generators will act on the right as in $S(\mu)$ by transforming $\dim S(\lambda)\times \dim S(\mu)$ different traceless tensors of shape $(\lambda,\mu)$ into each other. The  generators  $\mathsf{Q}_{i\bar{\imath}}$ and $\mathsf{Q}_{\bar{\imath}i+1}$ will act across the wall by contracting a covariant index with a contravariant one. In view of the tracelessness condition this action is trivial.
Thus, the space of all traceless tensors $t_0(\lambda,\mu)$ of the same shape $(\lambda,\mu)$ is isomorphic to a direct sum of $\dim t_0(\lambda,\mu)$ modules $\Delta_{f,f}(\lambda,\mu)\simeq S(\lambda)\times S(\mu)$ of the walled Brauer algebra.
A very important observation is the \emph{triviality of the centralizer} of a traceless tensor
\begin{equation}
 \End_{\gl(M|N)} t_0(\lambda,\mu)\simeq \mathbb{C}\ .
\label{eq:centralizer_simple}
\end{equation}
This is a generalization of the Schur lemma for $\gl(n)$ traceless tensors, which are irreducible. The statement~\eqref{eq:centralizer_simple} follows immediately from the action of the walled Brauer algebra in the space of traceless tensors of shape $(\lambda,\mu)$ that we have just described. 
It means that traceless tensors are a very special type of indecomposables, namely any $\gl(M|N)$ Casimir is diagonalizable and proportional to the identity in a traceless tensor. 
It should be noticed that this is typical of highest weight or Kac modules \cite{Kac77a, Kac77b}.   
\begin{assumption}{1}\label{ass:hw}
 Traceless tensors are highest weight modules.
\end{assumption}
\noindent This means that there is a Borel subalgebra $\mathfrak{b}$ of $\gl(M|N)\simeq \mathfrak{b}\oplus \mathfrak{n}^-$ and a $\mathfrak{b}$--highest weight vector $\mathbf{v}\in t_0(\lambda,\mu)$  such that the full tensor $t_0(\lambda,\mu)$ can be generated from $\mathbf{v}$ by repeated action of $\mathfrak{n}^-$. We shall see later how to choose $\mathfrak{b}$ for given $t_0(\lambda,\mu)$.

Consider now the vector space $\delta_{L,L}(\lambda,\mu)$ of all possible embeddings  of traceless tensors of shape $(\lambda,\mu)$ and ranks $(f,f)$ into the spin chain $(V\otimes V^*)^{\otimes L}$. It consists of tensor products of tensors $t_0(\lambda,\mu)$ with  $\gl(M|N)$--invariants of $(V\otimes V^*)^{\otimes (L-f)}$.
Notice that there is a unique $\gl(M|N)$ invariant in $V\otimes V^*$, which can be written as   $e_i\otimes e^i$.
Representing this invariant by an edge  connecting two vertices across the wall and the traceless tensors $t_0(\lambda,\mu)$ as we did before,
we can visualize $\delta_{L,L}(\lambda,\mu)$ as a diagram with $L$ vertices on each side of the wall and $L-f$ edges connecting pairs of vertices across the wall.
Thus, we reconstruct the same diagrammatic representation of $\delta_{L,L}(\lambda,\mu)$ as for $\Delta_{L,L}(\lambda,\mu)$. 
This proves that  all the relations between the generators of $B_{L,L}(n)$ satisfied in $\Delta_{L,L}(\lambda,\mu)$
will be satisfied in $\delta_{L,L}(\lambda,\mu)$ as well.
The converse is generally not true, meaning that $\delta_{L,L}(\lambda,\mu)$ is generally only a quotient of ($\dim t_0(\lambda,\mu)$ direct sums of) $\Delta_{L,L}(\lambda,\mu)$.
We stress that neither $\delta_{L,L}(\lambda,\mu)$ nor $\Delta_{L,L}(\lambda,\mu)$ are necessarily simple $B_{L,L}(n)$ modules and, therefore, the quotient can be non-trivial.

The ABA in some grading $\Sigma$ provides $\gl(M|N)$  Bethe eigenvectors of highest weight with respect to the Borel subalgebra $\mathfrak{b}_\Sigma$ determined by $\Sigma$ and the ordering~\eqref{eq:tot_ordering_bas_vecs}.
Therefore, in order to match the numerical spectrum of $\mathsf{H}$ in $\Delta_{L,L}(\lambda,\mu)$ with the exact spectrum of $H$ by the ABA we need to know the highest weight of a traceless tensor $t_0(\lambda,\mu)$ at least in one grading $\Sigma$. We explain below how to evaluate it.

Consider the Young diagrams corresponding to the shape $(\lambda,\mu)$ of a \emph{full} tensor $t(\lambda,\mu)$.
The basis vectors in the tensor subspace of co(ntra)variant shape $\lambda\, (\mu)$ can be represented by co(ntra)variant Young supertableaux of shape $\lambda\, (\mu)$, that is Young diagrams of shape $\lambda\, (\mu)$ with a fundamental weight $\epsilon_i\, (-\epsilon_j)$ inscribed in every box.
The pattern of weights within the Young diagrams must satisfy the supersymmetrization rules, that is i) in the same row the bosonic (fermionic) weights are weakly (strongly) ordered w.r.t. each other, ii) in the same column the bosonic (fermionic) weights are strongly (weakly) ordered w.r.t. each other
and iii) bosonic weights are weakly (strongly) ordered w.r.t. fermionic weight in the same row (column).
%The total ordering is given by ~(\ref{eq:tot_ordering_bas_vecs}, \ref{eq:dual_ord}).
The weight of a supertableau is the sum of all weights it carries in its boxes.

\begin{figure}%
\psfrag{mu}{$\mu$}
\psfrag{lam}{$\lambda$}
\psfrag{r1}{$r_1$}
\psfrag{r2}{$r_2$}
\psfrag{r3}{$r_3$}
\psfrag{r4}{$r_4$}
\psfrag{r5}{$r_5$}
\psfrag{x1}{$\bar{r}_1$}
\psfrag{x2}{$\bar{r}_2$}
\psfrag{x3}{$\bar{r}_3$}
\psfrag{x4}{$\bar{r}_4$}
\psfrag{x5}{$\bar{r}_5$}
\psfrag{c1}{$c_1$}
\psfrag{c2}{$c_2$}
\psfrag{c3}{$c_3$}
\psfrag{c4}{$c_4$}
\psfrag{y1}{$\bar{c}_1$}
\psfrag{y2}{$\bar{c}_2$}
\psfrag{y3}{$\bar{c}_3$}
\psfrag{y4}{$\bar{c}_4$}
\psfrag{e1}{$\epsilon_1$}
\psfrag{e2}{$\epsilon_2$}
\psfrag{e3}{$\epsilon_3$}
\psfrag{e4}{$\epsilon_4$}
\psfrag{e5}{$\epsilon_5$}
\psfrag{e6}{$\epsilon_6$}
\psfrag{e7}{$\epsilon_7$}
\psfrag{e8}{$\epsilon_8$}
\psfrag{e9}{$\epsilon_9$}
\psfrag{d1}{$-\epsilon_1$}
\psfrag{d2}{$-\epsilon_2$}
\psfrag{d3}{$-\epsilon_3$}
\psfrag{d4}{$-\epsilon_4$}
\psfrag{d5}{$-\epsilon_5$}
\psfrag{d6}{$-\epsilon_6$}
\psfrag{d7}{$-\epsilon_7$}
\psfrag{d8}{$-\epsilon_8$}
\psfrag{d9}{$-\epsilon_9$}
\psfrag{s}{$\Sigma$}
\psfrag{v}{$v_\Sigma(\lambda)$}
\psfrag{bv}{$\bar{v}_\Sigma(\mu)$}
\psfrag{bh}{$\bar{h}_\Sigma(\mu)$}
\psfrag{h}{$h_\Sigma(\lambda)$}
\centerline{\includegraphics[scale=1.3]{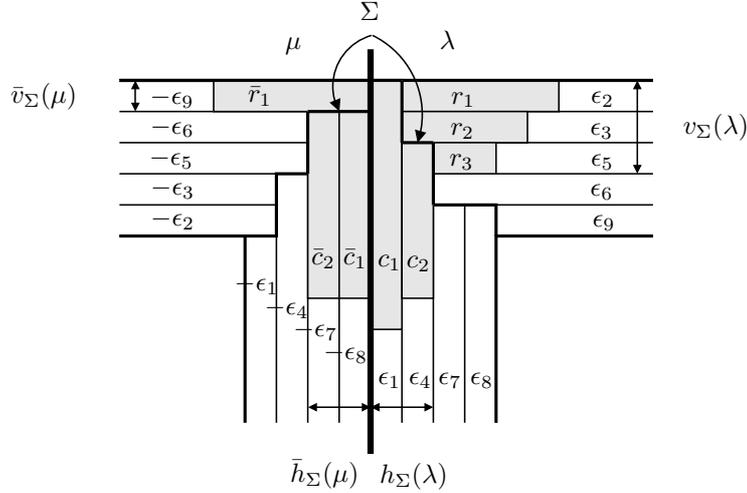}}
\caption{Covariant Young tableau $\lambda$ and contravariant Young tableau $\mu$ of $(5,4)$--hook shape for $\gl(5|4)$. Bosonic fundamental weights belong to $B=\{\epsilon_2,\epsilon_3,\epsilon_5,\epsilon_6,\epsilon_9\}$, while fermionic weights  to $F=\{\epsilon_1,\epsilon_4,\epsilon_7,\epsilon_8\}$. Correspondingly $-B=\{-\epsilon_9,-\epsilon_6,-\epsilon_5,-\epsilon_3,-\epsilon_2\}$ and $-F=\{-\epsilon_8,-\epsilon_7,-\epsilon_4,-\epsilon_1\}$. The values of $r_4$, $r_5$, $c_3$, $c_4$, $\bar{r}_2,\dots,\bar{r}_5$, $\bar{c}_3$, $\bar{c}_4$ are zero.}
\label{fig:young}
\end{figure}

%The highest weight of $t(\lambda,\mu)$ is the sum of the highest weight covariant Young supertableux of shape $\lambda$ and the highest weight contravariant Young supertableux of shape $\mu$.
Both $\lambda$ and $\mu$ must fit into a hook whose horizontal (vertical) arm is of width $M\, (N)$, otherwise $t(\lambda,\mu)$ vanishes identically because it is not possible to fill in the Young diagrams and get Young supertableaux compatible with the supersymmetrization rules.
The highest weight of a supertableau depends on the grading.
The choice of grading is a splitting of the set of basis vectors into two sets $B\, (F)=\{\epsilon_i\mid |i|\equiv 0\, (1)\}^{<}$ which are ordered w.r.t. the total ordering~\eqref{eq:tot_ordering_bas_vecs}.
Equivalently, it can be represented by paths connecting the two corners of the $(M,N)$--hooks as represented in fig.~\ref{fig:young}.
Fix these paths and consider Young diagrams $\lambda$, $\mu$
with rows $\lambda_i$, $\mu_i$ and columns $\lambda'_i$, $\mu'_i$.
Then, the highest weight of $t(\lambda,\mu)$ can be written in the following form
\begin{equation}\label{eq:weight_tensor}
 \Lambda_\Sigma(\lambda,\mu) = \sum_{i=1}^M [r_i\epsilon_{b(i)}-\bar{r}_i\epsilon_{\bar{b}(i)}]+\sum_{i=1}^N  [c_i\epsilon_{f(i)} -\bar{c}_i\epsilon_{\bar{f}(i)}]
\end{equation}
where $b(i)$ and $f(i)$ are the elements at position $i$ in the ordered sets $B$ and $F$,
 $\bar{b}(i)$ and $\bar{f}(i)$ are the elements at position $i$ in the ordered sets $-B$ and $-F$, 
 $r_i=\max(0,\, \lambda_i - \sum_{j=1}^{b(i)}(1-(-1)^{|i|})/2)$, $c_i=\max(0,\, \lambda'_i- \sum_{j=1}^i (1-(-1)^{|i|})/2$,
$\bar{r}_i = \max(0,\, \mu_i-\sum_{i=\bar{b}(i)}^M(1-(-1)^{|i|})/2 )$ and $\bar{c}_i = \max(0,\, \mu'_i-\sum_{i=\bar{f}(i)}^N(1+(-1)^{|i|})/2 )$
are number of boxes in a row or column of $\lambda$ or $\mu$ overpassing the grading paths as represented in fig.~\ref{fig:young}.

The highest weight component of a tensor $t(\lambda,\mu)$ w.r.t. the grading $\Sigma$ will belong to the traceless subspace $t_0(\lambda,\mu)$ if the highest weight Young supertableau of shape $(\lambda,\mu)$ does not contain some fundamental weight $\pm \epsilon_i$ in both $\lambda$ and $\mu$.
Otherwise, the corresponding highest weight component of $t(\lambda,\mu)$ w.r.t. the grading $\Sigma$ will either i) not belong to $t_0(\lambda,\mu)$ or ii) generate a submodule of $t_0(\lambda,\mu)$ isomorphic to the embedding of a quotient of a lower rank tensor $t_0(\lambda',\mu')$. If the latter case holds, then the possible Young diagrams $(\lambda',\mu')$ are obtained from the Young diagrams $(\lambda,\mu)$ by removing pairs of boxes from the highest weight Young supertableau of shape $(\lambda,\mu)$: a box of $\lambda$ carrying some weights $\epsilon_i$ together with a box of $\mu$ carrying the opposite weight $-\epsilon_i$.
Moreover, in the case ii) the highest weight of the top~\footnote{The top of a module is the quotient by the intersection of all maximal ideals. The top of a Kac module is the irreducible representation of highest weight.} of $t_0(\lambda,\mu)$ will be smaller then the highest weight of the submodule $t_0(\lambda',\mu')$.
Therefore, $t_0(\lambda,\mu)$ cannot be a Kac module w.r.t. $\mathfrak{b}_\Sigma$.

We call $\gl(M|N)$--\emph{admissible} the shapes $(\lambda,\mu)$ for which the $\gl(M|N)$ traceless tensors $t_0(\lambda,\mu)$ neither vanish nor are isomorphic to lower rank traceless tensors.
We are now ready to answer the very important question: what are the admissible shapes of traceless tensors?
According to the previous discussion on the highest weight component of a tensor $t(\lambda,\mu)$, a shape $(\lambda,\mu)$ is admissible if there \emph{exists} a grading $\Sigma$ such that 
the highest weight Young supertableau of shape $(\lambda,\mu)$ does not contain any fundamental weight $\pm \epsilon_i$ both in $\lambda$ and in $\mu$.
Consequently, for a traceless tensor $t_0(\lambda,\mu)$ of admissible shape there is a grading $\Sigma$ and a corresponding highest weight~\eqref{eq:weight_tensor} such that  no cancellation between the $r_i$ and $\bar{r}_i$ or $c_i$ and $\bar{c}_i$ terms can occur.
If $v_\Sigma(\lambda)$, $\bar{v}_\Sigma(\mu)$ denote the number of nonzero ``reduced rows'' $r_i$, $\bar{r}_i$ and $h_\Sigma(\lambda)$, $\bar{h}_\Sigma(\mu)$ denote the number of non-zero ``reduced'' columns
$c_i$, $\bar{c}_i$, then one must have
\begin{equation}
\exists \Sigma \quad \text{such that} \quad v_\Sigma(\lambda)+\bar{v}_\Sigma(\mu)\leq M,\qquad h_\Sigma(\lambda)+\bar{h}_\Sigma(\mu)\leq N
\label{eq:restr_form}
\end{equation}
for a $\gl(M|N)$--admissible shape $(\lambda,\mu)$, as represented in fig.~\ref{fig:young}
These admissible shapes nicely reduce to hook shapes for purely covariant or contravariant $\gl(M|N)$ tensors and to staircases for $\gl(n)$ traceless tensors \cite{staircase}.

We say that a shape $(\lambda,\mu)$ of a traceless tensor $t_0(\lambda,\mu)$ is $\Sigma$--admissible if the inequalities in eq.~\eqref{eq:restr_form} are satisfied.
Let $K_\Sigma(\Lambda)$ be the Kac module of highest weight $\Lambda$ w.r.t. $\mathfrak{b}_\Sigma$. 
We can  make now assumption~\ref{ass:hw} more precise.
\begin{assumption}{1$'$}\label{ass:hw2}
 The following isomorphism holds for $\Sigma$--admissible shapes $(\lambda,\mu)$
\begin{equation*}
 t_0(\lambda,\mu)\simeq K_\Sigma(\Lambda_\Sigma(\lambda,\mu))\ .
\end{equation*}
\end{assumption}
\noindent 
This assumption in combination with the general theory of Kac modules \cite{Kac77a, Kac77b} is very useful for counting highest weight vectors. Namely, if $(\lambda,\mu)$ is $\Sigma$--admissible then the number of highest weight vectors in $t_0(\lambda,\mu)$ w.r.t. $\mathfrak{b}_\Sigma$ is equal to the number of irreducible subquotients.
Noticing that a highest weight vector cannot generate more then a highest weight module, we prove the following claim in app.~\ref{sec:proofs}.
\begin{claim}\label{claim:cv}
Highest weight vectors of $(V\otimes V^*)^{\otimes L}$ w.r.t. any Borel subalgebra belong to submodules isomorphic to traceless tensors $t_0(\lambda,\mu)$, $\lambda,\mu\vdash f=0,1,\dots,L$ of admissible shape.
\end{claim}

To sum up, we have explained the connexion between traceless tensors $t_0(\lambda,\mu)$ of admissible shapes $(\lambda,\mu)$ and standard modules $\Delta_{L,L}(\lambda,\mu)$. Secondly, if $(\lambda,\mu)$ is $\Sigma$-admissible, then eq.~\eqref{eq:weight_tensor} allows to compute the highest weight of $t_0(\lambda,\mu)$ w.r.t. the Borel subalgebra $\mathfrak{b}_\Sigma$. Evaluating the highest weight of $t_0(\lambda,\mu)$ w.r.t. arbitrary Borel subalgebras is more delicate, mainly because of indecomposability issues.
Finally, claim~\ref{claim:cv} indicates a natural relationship between traceless tensors and Bethe vectors, which have the highest weight property, constructed in the framework of ABA.

%%%%%%%%%%%%%%%%%%%%%%%%%%%%%%%%%%%%%%%%%%%%%%%%%%%%%%%%%
%%%%%%%%%%%%%%%%%%%%%%%%%%%%%%%%%%%%%%%%%%%%%%%%%%%%%%%%%
\subsection{Spectrum}
\label{sec:spec}
%%%%%%%%%%%%%%%%%%%%%%%%%%%%%%%%%%%%%%%%%%%%%%%%%%%%%%%%%
%%%%%%%%%%%%%%%%%%%%%%%%%%%%%%%%%%%%%%%%%%%%%%%%%%%%%%%%%

We present the spectrum of $\mathsf{H}$ in various $\Delta_{L,L}(\lambda,\mu)$ standard modules of $B_{L,L}(n)$ in tab.~\ref{tab:1}.
%%%%%%%%%%%%%%%%%%%%%%%
\begin{table}[t]\label{tab:1}
\begin{center}
\caption{Lowest eigenvalue of $\mathsf{H}$ in $\Delta_L(\lambda,\mu)$ with opposite sign.}
\begin{tabular}{|c|c|ccccccc|}
\hline
 $L$& $n$ & $(1^2,2)$ & $(2,2)$ & $(1^3,21)$ & $(1^3,3)$ & $(21,21)$ & $(21,3)$ & $(3,3)$\\ \hline
 \multirow{8}{*}{5} & 1 &
$32.130$   & $29.797$ & $29.461$ & $27.453$ & $27.156$  & $24.101$ & $21.446$  \\ 
&  &
$\pm0.591 i$&            &         &           &          &             &  \\ 
& 2 &
$23.237$    & $21.904$ & $22.626$  & $20.498$ & $20.506$    & $17.999$   & $15.548$ \\ 
&  &
$\pm 0.339i$&           &           &           &          &          & \\
& 3 &
$20.763$    & $20.130$ & $20.554$ & $18.530$ & $18.762$  & $16.572$      &  $14.472$ \\
&   &
              &           &          &          &        &       &  \\
& 4 &
$19.968$ & $19.487$ & $19.588$ & $17.682$ & $18.139$   &$16.180$  &  $14.153$ \\
&  &
           &            &          &           &           &             &  \\ \hline
 \multirow{8}{*}{6} & 1 &
41.558 & 39.639 & 38.640 &        36.835 & 36.559 & 33.762&  31.264  \\ 
&  &
             &       &  $\pm 0.355 i$ &  & $\pm 0.513 i$&   $\pm 0.127i$ &  \\ 
& 2 & 29.537  & 28.282 & 28.802  & 26.976 & 26.984 & 24.784& 22.607 \\ 
&  &          &           & $\pm 0.267 i$ &       & $\pm 0.426 i$  &  $\pm 0.148i$  & \\
& 3 & 26.123 & 25.600 & 25.884  & 24.210 & 24.365&  &  20.713\\
&  &          &         & $\pm 0.194i$ &     &  $\pm0.258i$& 22.550  &  \\
& 4 & 24.917 & 24.557 & 24.566 & 23.031 &23.416&  21.800 &  20.021\\
&  &         &            &  $\pm 0.129i$ &      &   &  &  \\ \hline

\end{tabular}
\end{center}
\end{table}
%%%%%%%%%%%%%%%
At a first glance, it appears that the vacuum always lies in $\Delta_{L,L}(\emptyset,\emptyset)$.
To check more thoroughly this vacuum hypothesis we need an additional assumption on the spectrum.
Consider the spectral sets of $\mathsf{H}$ defined as
\begin{equation*}
\spec f=\bigcup_{\lambda,\mu\vdash f } \spec\Delta_{L,L}(\lambda,\mu)\ .
\end{equation*}
Notice from tab.~\ref{tab:1} that the lowest eigenvalue in $\spec f$ always lies in $\Delta_{L,L}(1^f,1^f)$, where $1^f$ denotes the Young diagram with a single column of length $f$. We have checked this observation extensively \textbf{(more details)}.
\begin{assumption}{2}\label{ass:ord}
 The lowest eigenvalues of $\mathsf{H}$ in $\spec f$ always lies in $\Delta_{L,L}(1^f,1^f)$.
\end{assumption}
\noindent Comparing only the lowest eigenvalues in $\Delta_{L,L}(1^f,1^f)$ allows us to gain several spin chain length units and check the vacuum hypothesis further, see tab.~\ref{tab:2}.

%%%%%%%%%%%%%%%%%%%%%%%
\begin{table}\label{tab:2}
\begin{center}
\caption{Lowest eigenvalue of $\mathsf{H}$ in $\Delta_L(1^k,1^k)$ with opposite sign.}
\begin{tabular}{|c|c|ccccccccc|}
\hline
 $L$& $n$ & $k=0$ & $k=1$ & $k=2$ & $k=3$ & $k=4$ & $k=5$ & $k=6$ & $k=7$ & $k=8$\\ \hline
 \multirow{4}{*}{5} & 1 & 40.000 & 38.846 & 36.564 & 32.710 & 26.385& 20.000 & & & \\ 
& 2 & 28.062 & 26.369 & 26.156 & 25.082 & 22.606& 20.000 & &  &\\ 
& 3 & 24.625 & 22.950 & 22.944 & 22.640 & 21.407& 20.000 &  &  & \\
& 4 & 23.123 & 21.631 & 21.456 & 21.456 & 20.828& 20.000 &  &  & \\ \hline
\multirow{4}{*}{6} & 1 & 48.000 & 46.723 & 45.626 & 41.538 & 36.782& 30.397 & 24.000 &  & \\ 
& 2 & 33.550 & 32.126 & 32.126 & 30.902 & 29.253& 26.647 & 24.000 &  & \\ 
& 3 & 29.388 & 27.989 & 28.024 & 27.617 & 26.885& 25.476 & 24.000 &  & \\ 
& 4 & 27.574 & 26.339 & 26.146 & 26.086 & 25.754& 24.921 & 24.000 &  & \\ \hline
 \multirow{4}{*}{7} & 1 & 56.000 & 55.134 & 53.631 & 50.894 & 46.015 & 40.796 & 34.399 & 28.000 &\\ 
& 2 & 39.054 & 37.826 & 37.699 & 36.992 & 35.260 & 33.307 & 30.660 & 28.000 & \\ 
& 3 & 34.172 & 32.971 & 32.939 & 32.748 & 31.954& 30.981 & 29.504 & 28.000 & \\
& 4 & 32.046 & 30.991 & 30.800 & 30.800 & 30.420 & 29.885 & 28.962 & 28.000 & \\ \hline
\multirow{4}{*}{8} & 1 & 64.000 & 63.035 & 63.296 & 59.127 & 55.589 & 50.296  & 44.799 & 38.400 & 32.000\\ 
& 2 & 44.569 & 43.488 & 43.488 & 42.533 & 41.457 & 39.453 & 37.325 & 34.664 & 32.000\\ 
& 3 & 38.970 & 37.917 & 39.480 & 37.652 & 37.146& 36.134 & 35.019 & 33.515 & 32.000\\ 
& 4 &  36.530    & 35.611 & 35.428 & 35.391 & 35.169& 34.603 & 33.944 & 32.982 & 32.000\\ \hline
\end{tabular}
\end{center}
\end{table}
%%%%%%%%%%%%%%%

To extract the spectrum of the $\gl(n+N|N)$ spin chain Hamiltonian~\eqref{eq:ham} from the spectrum of the algebraic Hamiltonian $\mathsf{H}$ we do the following.
For every pair of Young diagrams $(\lambda,\mu)$ with $f=0,1,\dots, L$ boxes and of admissible shape we pick a grading $\Sigma$ such that the inequalities in eq.~\eqref{eq:restr_form} are satisfied.
Then we try to reproduce the spectrum of $\mathsf{H}$ in $\Delta_{L,L}(\lambda,\mu)$ by means of eq.~\eqref{eq:spec_ham_mom} from numerical solutions of BAE~\eqref{eq:BAE_short} in the the form determined by the grading $\Sigma$ and for root numbers corresponding
to the highest weight $\Lambda_\Sigma(\lambda,\mu)$ in eqs.~(\ref{eq:weight_BV1}, \ref{eq:weight_tensor}).~\footnote{One could work with a single form of BAE, say, that corresponding to the distinguished gradation $\Sigma_0$.
However, the weight of Bethe vectors  reproducing eigenvalues of $\mathsf{H}$ in $\Delta_{L,L}(\lambda,\mu)$ will no longer be given by eq.~\eqref{eq:weight_tensor} if the shapes of $(\lambda,\mu)$ do not satisfy the inequalities~\eqref{eq:restr_form} w.r.t. $\Sigma_0$.}
Not all eigenvalues can be reproduced in this way. This happens because, as we have explained in sec.~\ref{sec:tens}, 
the vector space $\delta_{L,L}(\lambda,\mu)$ of all possible embeddings of $\gl(n+N|N)$ traceless tensors  $t_0(\lambda,\mu)$ into $(V\otimes V^*)^{\otimes L}$ can be identified with only a quotient of the standard module $\Delta_{L,L}(\lambda,\mu)$ of $B_{L,L}(n)$.
When an eigenvalue of $\mathsf{H}$ can be reproduced this way, we assume that a corresponding non-vanishing Bethe vector exist. Otherwise, we assume that there is no eigenstate of $H$ corresponding to that eigenvalue. So, we rely on the completeness of the ABA, at least as far as the spectrum is concerned.

Denote the spectrum of the $\gl(M|N)$ spin chain Hamiltonian~\eqref{eq:ham} by $\spec H_{M|N}$.
The central idea of this section was the existence of an abstract algebra $B_{L,L}(n)$ such that the centralizers of the series $N=0,1,2,\dots$ of $\gl(n+N|N)$ spin chains $\mathcal{C}(L)$ provide different, $N$ dependent, representations of $B_{L,L}(n)$.
This suggests that the intersection $\cap_{N\in \mathbb{Z}^+}\spec H_{n+N|N}$ might be non-trivial.
In fact, with the cohomological techniques developed in \cite{Candu:2010yg} one can prove the following relationship between the spectral sets $\spec H_{N+n|N}$ with $n$ fixed
\begin{equation}
\spec H_{n|0}\subset \spec H_{n+1|1}\subset \spec H_{n+2|2}\subset \cdots\subset \mathsf{H}\ . 
\label{eq:embed_spec}
\end{equation}
This ``embedding of spectra'' is a very interesting and general feature of supergroup spin chains and one might wonder how does it carry on to the field theory description of the continuum limit.
In this respect, two scenarios are possible.
The first possibility is that $\spec H_{n+N'|N'}$ becomes a very excited subset within $\spec H_{n+N''|N''}$, where $N'<N''$, and decouples from it in the thermodynamic limit $L\to \infty$.
This means that the vacuum energy per site of $H_{n+N'|N'}$ is higher then the vacuum energy per site of $H_{n+N''|N''}$ in the thermodynamic limit.
The second, more interesting, possibility is that the vacuum energies per site for both Hamiltonians coincide in the thermodynamic limit.
The first possibility occurs, for instance, in the $V^{\otimes L}$ chains with Hamiltonian $\pm \sum P_{i,i+1}$.

In next section we describe the mechanism which is behind the embedding of spectra~\eqref{eq:embed_spec}  at the level of BAE.
The answer to the question of how excited $\spec H_{n+N'|N'}$ is within $\spec H_{n+N''|N''}$ for $N'< N''$ will have to wait until sec.~\ref{sec:rl}.

%%%%%%%%%%%%%%%%%%%%%%%%%%%%%%%%%%%%%%%%%%%%%%%%%%%%%%%
%%%%%%%%%%%%%%%%%%%%%%%%%%%%%%%%%%%%%%%%%%%%%%%%%%%%%%%
\section{Restriction and lift of BAE}
\label{sec:rl}
%%%%%%%%%%%%%%%%%%%%%%%%%%%%%%%%%%%%%%%%%%%%%%%%%%%%%%%
%%%%%%%%%%%%%%%%%%%%%%%%%%%%%%%%%%%%%%%%%%%%%%%%%%%%%%%

At the end of the previous section we have explained the phenomenon~\eqref{eq:embed_spec} of embedding of spectra. For it to work, it was important that there is a series  $\{\gl(n+N|N)\}_{N\in\mathbb{Z}^+}$ of spin chains  endowed with Hamiltonians $H_{n+N|N}$ which are all different representations of the same algebraic Hamiltonian $\mathsf{H}\in B_{L,L}(n)$.~\footnote{Strictly speaking, it does not make sense to talk about different representations of a single matrix $\mathsf{H}$. What we mean here is that the $H_{n+N|N}$ are the image of $\mathsf{H}$ in different representations of $B_{L,L}(n)$.}
The integrability did not matter.
In this section we would like to understand how this curious phenomenon arises at the level of BAE describing the spectra of very general $\gl(M|N)$ integrable Hamiltonians and what structure is responsible for it.

Fix a grading $\Sigma = \{\sigma_i=(-1)^{|i|}\}_{i=1}^{M+N}$  
of the ordered basis vectors~\eqref{eq:tot_ordering_bas_vecs} of the $\gl(M|N)$ fundamental representation $V$ and let $\Delta^{M|N}_0=\{\alpha_j=\epsilon_j-\epsilon_{j+1}\}_{j=1}^{r}$ be the corresponding simple root system, where $r=M+N-1$ is the rank.
We have explained how to construct the ABA in the grading $\Sigma$ for the spin chain $(V\otimes V^*)^{\otimes L}$ in sec.~\ref{sec:formalism}.
The Bethe vectors are highest weight vectors of weight~\eqref{eq:weight_BV1} w.r.t. the Borel subalgebra $\mathfrak{b}_\Sigma$ determined by the grading $\Sigma$ and the ordering~\eqref{eq:tot_ordering_bas_vecs}, that is by $\Delta^{M|N}_0$.
We shall restrict to Bethe vectors whose weights are given by highest weight Young supertableaux of $\Sigma$--admissible shapes $(\lambda,\mu)$ according to eq.~\eqref{eq:weight_tensor}.
Due to fundamental r\^ole of traceless tensors explained in sec.~\ref{sec:tens} and claim~\ref{claim:cv}, it is clear that with the imposed restriction, one must consider the ABA in all the gradings in order to recover the full spectrum of the spin chain Hamiltonian $H$.
For a fixed grading and a corresponding simple root system, fig.~\ref{fig:young} and eq.~\eqref{eq:weight_tensor} implies that we are restricting to Bethe vectors of weight $w=\sum_{i=1}^{M+N}w^i \epsilon_i$, $w^i\in\mathbb{Z}$ such that for every simple bosonic root $\alpha_j=\epsilon_j-\epsilon_{j+1}$ one has $w^i \geq w^{i+1}$,
while for every simple fermionic  root one has the following implications
\begin{align}\label{eq:cond_TL}
 w^{j\phantom{+1}}<0 &\longrightarrow w^{j+1}<0\ ,& w^{j\phantom{+1}} =0 &\longrightarrow w^{j+1}\leq 0\\ \notag
 w^{j+1}>0 &\longrightarrow w^{j\phantom{+1}}>0\ ,&  w^{j+1}=0&\longrightarrow w^{j\phantom{+1}} \geq 0 \ .
\end{align}
The main reason for introducing this restrictions and working with BAE in multiple gradings is the bounds on the number of Bethe roots resulting from eqs.~\eqref{eq:cond_TL}.

\subsection{Restriction}

We wish to consider the BAE in the form~\eqref{eq:BAE_Qform} corresponding to a simple root $\alpha_k$ such that the following assumptions hold
\begin{description}
	\item[A1] $\alpha_k$ is odd, that is $\sigma_k \sigma_{k+1}=-1$
	\item[A2] $\alpha_k$ has no source terms or, equivalently, $k\neq 1, r$
\end{description}
With these assumptions, we have $\nu^{(k)}$ BAE for $\alpha_{k-1}$ of the form
\begin{equation}\label{eq:red1} \frac{\sigma_{k-1}\Lambda_{k-1}(u^{(k-1)}_j)}{\sigma_{k}\Lambda_{k}(u^{(k-1)}_j)}=- \frac{Q_{k-2}(u^{(k-1)}_j)Q_{k-1}(u^{(k-1)}_j+\sigma_k)Q_{k}(u^{(k-1)}_j-\sigma_k)}{Q_{k-2}(u^{(k-1)}_j+\sigma_{k-1})Q_{k-1}(u^{(k-1)}_j-\sigma_{k-1})Q_{k}(u^{(k-1)}_j)}\ ,
\end{equation}
$\nu^{(k)}$ equations for $\alpha_k$
\begin{equation}\label{eq:red2} 1=\frac{Q_{k-1}(u^{(k)}_j)Q_{k+1}(u^{(k)}_j+\sigma_k)}{Q_{k-1}(u^{(k)}_j+\sigma_k)Q_{k+1}(u^{(k)}_j)}
\end{equation}
and $\nu^{(k+1)}$ equations for $\alpha_{k+1}$
\begin{equation}\label{eq:red3} \frac{\sigma_k \Lambda_{k+1}(u^{(k+1)}_j)}{\sigma_{k+2}\Lambda_{k+2}(u^{(k+1)}_j)}= \frac{Q_{k}(u^{(k+1)}_j)Q_{k+1}(u^{(k+1)}_j+\sigma_{k+2})Q_{k+2}(u^{(k+1)}_j-\sigma_{k+2})}{Q_{k}(u^{(k+1)}_j-\sigma_k)Q_{k+1}(u^{(k+1)}_j+\sigma_k)Q_{k+2}(u^{(k+1)}_j)}\ .
\end{equation}
Notice that if $\nu^{(k-1)}=\nu^{(k+1)}$ then one can reduce the
BAE~\eqref{eq:BAE_Qform} for the  $\gl(M|N)$ spin chain $(V\otimes V^*)^{\otimes L}$ to the BAE for the $\gl(M-1|N-1)$ spin chain of the same type $(V\otimes V^*)^{\otimes L}$ by
\begin{description}
	\item[R1] identifying the Bethe roots corresponding to $\alpha_{k-1}$ and $\alpha_{k+1}$
\begin{equation}  \{u^{(k-1)}_j\}_{j=1}^{\nu^{(k-1)}}=\{u^{(k+1)}_j\}_{j=1}^{\nu^{(k+1)}}
\end{equation}
\item[R2] multiplying the BAE for $\alpha_{k-1}$ and $\alpha_{k+1}$ corresponding to, say $u^{(k-1)}_j=u^{(k+1)}_j$
\begin{equation}\label{eq:red}
 \frac{\sigma_{k-1}\Lambda_{k-1}(u^{(k- 1)}_j)}{\sigma_{k+2}\Lambda_{k+2}(u^{(k- 1)}_j)}=-
 \frac{Q_{k-2}(u^{(k- 1)}_j)Q_{k- 1}(u^{(k- 1)}_j+\sigma_{k+2})Q_{k+2}(u^{(k- 1)}_j-\sigma_{k+2})}{Q_{k-2}(u^{(k-1)}_j+\sigma_{k-1})Q_{k- 1}(u^{(k- 1)}_j-\sigma_{k-1})Q_{k+2}(u^{(k- 1)}_j)}\ .
\end{equation}
\end{description}
Indeed,  eq.~\eqref{eq:red2} is trivially satisfied  because according to \textbf{R1} one has $Q_{k-1}(u)=Q_{k+1}(u)$. 
Furthermore, multiplying BAE according to \textbf{R2}, the Bethe roots $\{u^{(k)}_j\}_{j=1}^{\nu^{(k)}}$ drop off yielding a BAE of the form~\eqref{eq:red1} with $k$ replaced by $k+2$.
We call a \emph{restriction} of BAE the procedure \textbf{R1}--\textbf{R2}.

The restriction procedure can be given an algebraic meaning and, therefore, partially explained as follows. 
As described in detail in~\cite{Candu:2010yg}, choose
\begin{equation}\label{eq:q_choice}
Q=E_{k+1k} 
\end{equation}
to be the odd $\gl(M|N)$ element that squares to zero and defines the $\gl(M-1|N-1)$ chain $(V_{M-1|N-1}\otimes V^*_{M-1|N-1})^{\otimes L}$ as the $Q$--cohomology of the  $\gl(M|N)$  chain $(V_{M|N}\otimes V^*_{M|N})^{\otimes L}$.
A necessary condition for a highest weight vector $\omega\in (V_{M|N}\otimes V^*_{M|N})^{\otimes L}$ to yield a non-trivial $Q$--cohomology is for it to be  in the kernel of $Q$ and, therefore, one must have
\begin{equation}\label{eq:wt_ker}
[E_{kk+1},E_{k+1k}]\omega =0\quad \Rightarrow\quad \langle \wt(\omega),\alpha_k\rangle=0\ .
\end{equation}
Applying this constraint to a Bethe vector of weight~\eqref{eq:weight_BV1} one recovers the condition $\nu^{(k-1)}=\nu^{(k+1)}$ necessary for \textbf{R1} to hold.
The form of the reduced $\gl(M-1|N-1)$ BAE can also be easily understood.
The subquotient $ \big(\ker_{\langle\alpha_k,-\rangle} \sum_{i=1}^{M+N}\mathbb{C}\epsilon_i\big)/\mathbb{C}\alpha_k$ of the $\gl(M|N)$ weight space can be straightforwardly identified with the $\gl(M-1|N-1)$ weight space $\sum_{i\neq k,k+1}^{M+N}\mathbb{C}\epsilon_i$.
Therefore, the subquotient $ \big(\ker_{\langle\alpha_k,-\rangle} \Delta_0^{M|N}\big) /\mathbb{C}\alpha_k$ of the $\gl(M|N)$ simple root system $\Delta^{M|N}_0$ induced by cohomological reduction can be identified with a $\gl(M-1|N-1)$ simple root system $\Delta_0^{M-1|N-1}=  \{\alpha_1,\dots,\alpha_{k-2},\epsilon_{k-1}-\epsilon_{k+2},\alpha_{k+2},\dots,\alpha_r\}$.
The reduced BAE have the form~\eqref{eq:BAE_short} corresponding to precisely the simple root system $\Delta_0^{M-1|N-1}$.

\subsection{Lift}

\begin{figure}%
\psfrag{+}{$+$}
\psfrag{-}{$-$}
\psfrag{a1}{$\alpha_{k-1}$}
\psfrag{a2}{$\alpha_{k}$}
\psfrag{a3}{$\alpha_{k+1}$}
\psfrag{a4}{$\alpha_{k-1}\simeq\alpha_{k+1}$}
\psfrag{b1}{$\gl(M|N)$}
\psfrag{b2}{$\gl(M-1|N-1)$}
\psfrag{P}{$P$}
\psfrag{p}{$p$}
\centerline{\includegraphics[scale=1]{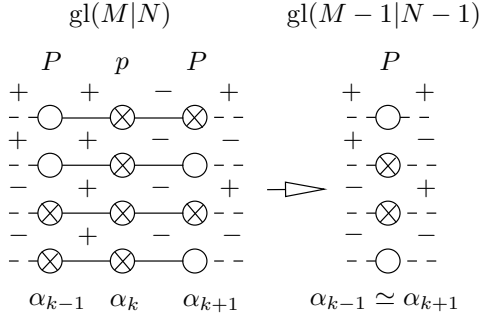}}
\caption{Restriction of BAE for supergroups. The dashed lines indicate possible roots on the left or on the right. For a root $\alpha_j=\epsilon_j-\epsilon_{j+1}$ with $j=k,k\pm 1$, we have indicated with $\pm$ signs the gradings of fundamental weight $\epsilon_j$ determining it.}
\label{fig:col}
\end{figure}

To resume, assuming \textbf{A1}--\textbf{A2} holds for the BAE of the spin chain $(V_{M|N}\otimes V^*_{M|N})^{\otimes L}$ written w.r.t. a simple root system $\Delta^{M|N}_0$, we showed that one can restrict them to the system of BAE for the spin chain $(V_{M-1|N-1}\otimes V^*_{M-1|N-1})^{\otimes L}$ written w.r.t. the simple root system $\Delta_0^{M-1|N-1}=\{\alpha_1,\dots,\alpha_{k-2},\epsilon_{k-1}-\epsilon_{k+2}, \alpha_{k+2},\dots,\alpha_r\}$ induced by  cohomological reduction.
This restriction is represented at the level of Dynkin diagrams in fig.~\ref{fig:col}.

Notice that we have not imposed any condition on the roots $u_j^{(k)}$.
Therefore, it is legitimate to ask if the BAE satisfying \textbf{A1}--\textbf{A2} actually admit solutions of type \textbf{R1}.
So, we are given a simple root system $\Delta_0^{M|N}$ and a solution $\{u_j^{(l)}\}_{j=1}^{\nu^{(l)}}$, $l=1,\dots,k-1,k+2,\dots,r$ of the 
$(V_{M-1|N-1}\otimes V^*_{M-1|N-1})^{\otimes L}$ BAE w.r.t the  simple root system $\Delta_0^{M-1|N-1}$ induced by cohomological reduction from $\Delta_0^{M|N}$ with $Q$ as in eq.~\eqref{eq:q_choice}.
We must show that there is a solution $\{u_j^{(l)}\}_{j=1}^{\nu^{(l)}}$, $l=1,\dots,r$ of the $(V_{M|N}\otimes V^*_{M|N})^{\otimes L}$ BAE w.r.t. $\Delta_0^{M|N}$ which restricts to the given one.
First of all, eq.~\eqref{eq:wt_ker} implies $\nu^{(k+1)}=\nu^{(k-1)}=P$. After defining $u^{(k+1)}_j=u^{(k-1)}_j$ for $j=1,\dots,P$ the task is reduced to finding a solution $\{u^{(k)}_j\}_{j=1}^p$ to either eq.~\eqref{eq:red1} or eq.~\eqref{eq:red3}, where we have set $p=\nu^{(k)}$.
If $p\geq P$, such a solution obviously exists. More then that, for $p>P$ there is a continuum of such solutions!
However, recall that for a fixed root system and corresponding BAE we have restricted to Bethe vectors such that their weights satisfy the constraints~\eqref{eq:cond_TL}.
If the solution we are looking for exists then the weight  of the corresponding Bethe vector $\omega$ can be written as $\wt(\omega)=\dots+(P-p)(\epsilon_k-\epsilon_{k+1})+\dots$.
The constraints~\eqref{eq:cond_TL} imply $p\leq P$.
So, for $p=P$ the solution \emph{always exists}.
In our numerical investigations we have observed that solutions might exist even for $p<P$.
We call the solutions of $\gl(M|N)$ BAE  constructed in this way from solutions of $\gl(M-1|N-1)$ BAE \emph{lifted solutions}.
We discard lifted solutions with $p< P$ for the following reason.
\begin{claim}\label{claim:coh}
 If a solution of $\gl(M-1|N-1)$ BAE admits lifts to solutions of $\gl(M|N)$ BAE corresponding to Bethe vectors of different weights, then all of them have zero $Q$--cohomology.
\end{claim}
\noindent We prove the claim in appendix~\ref{sec:B}.

Next, we show that \emph{the lift is unique}.
So, we have seen that lifted Bethe vectors correspond to lifted solutions with  $\nu^{(k)}=\nu^{(k\pm 1)}=P$.
Let us now prove that the $P$ equations, say,~\eqref{eq:red1} regarded as a constraint on the unknowns $\{u^{(k)}_j\}_{j=1}^P$ admit a unique solution. We start by multiplying both sides with $Q_k(u_j^{(k-1)})$ and then expand in  $\{u^{(k)}_j\}_{j=1}^P$.
What we get is a system of $P$ linear equations for the $P$ unknowns
\begin{equation*}
 s_l = \sum_{i_1<\cdots<i_l}u^{(k)}_{i_1}\dots u^{(k)}_{i_l}\ .
\label{eq:sym}
\end{equation*} 
Thus, for a  given solution $\{u_j^{(1)}\}_{j=1}^{\nu^{(j)}}$, $j=1,\dots,k-1,k+2,\dots,r$ of the $\gl(M-1|N-1)$ BAE, this system of linear equations admits a single solution for the $\{s_j\}_{j=1}^P$.
Finally, notice that the latter determine the set $\{u^{(k)}_j\}_{j=1}^P$ uniquely, because every Bethe root in this set is a solution of the polynomial equation of degree $P$
\begin{equation*}
\prod_{j=1}^P(u-u_j^{(k)}) = u^P +s_1 u^{P-1}+\dots- (-1)^P s_{P-1}u + (-1)^Ps_P=0\ .
\end{equation*}

There are important cases when one can compute the roots $\{u^{(k)}_j\}_{j=1}^P$ explicitly in terms of the other roots.
This happens if
\begin{description}
\item[C 1.1)] $\alpha_{k\pm 1}$ is odd and sourceless
\item[C 1.2)]  $\nu^{(k\pm 2)}=P$, 
\end{description}
where we meant that either we choose the plus sign or the minus sign overall. 
In the case \textbf{C~1} eq.~\eqref{eq:red1} or eq.~\eqref{eq:red3} can be rewritten in terms of variables~\eqref{eq:shifts} as
\begin{equation*}
\prod_{i=1}^P e_1(x^{(k\pm 1)}_j-x^{(k)}_i) =  \prod_{i=1}^P e_1(x^{(k\pm 1)}_j-x^{(k\pm 2)}_i)\ ,\quad j =1,\dots,P\ .
%\label{eq:}
\end{equation*}
We immediately read off the obvious solution
\begin{equation} \label{eq:freezing1}
 \{x^{(k)}_{j}\}_{j=1}^P = \{x^{(k\pm 2)}_j\}_{j=1}^P\ ,
\end{equation}
which we already know is unique.
Another important case is
\begin{description}
\item[C 2.1)] $k=2$, $\alpha_1$ is odd and $P=L$ 
\item[C 2.2)] $k=r-1$, $\alpha_r$ is odd and $P=L$.
\end{description}
In the case \textbf{C~2.1} the one has
\begin{equation*}
\prod_{i=1}^L e_1(x^{(1)}_j-x^{(2)}_i) =  \prod_{a=1}^L e_1(x^{(1)}_j-y_a)\ ,\quad j =1,\dots,L\ ,
%\label{eq:}
\end{equation*}
while in the case \textbf{C~2.2}
\begin{equation*}
\prod_{i=1}^L e_1(x^{(r)}_j-x^{(r-1)}_i) =  \prod_{a=1}^L e_1(x^{(r)}_j-\bar{y}_a)\ ,\quad j =1,\dots,L\ ,
%\label{eq:}
\end{equation*}
where we have used eqs.~\eqref{eq:BAE_short} with non-zero arbitrary inhomogeneities.
Again, we  read off the unique solutions
\begin{equation} \label{eq:freezing2}
 \{x^{(2)}_j\}_{j=1}^L=\{y_a\}_{a=1}^L\ ,\qquad
 \{x^{(r-1)}_j\}_{j=1}^L=\{y_{\bar{a}}\}_{a=1}^L\ .
\end{equation}
It is important to notice that inhomogeneities are essential to lift the degeneracy of solutions in the case \textbf{C~2}.

\subsection{Generalizations}

We have argued that the restriction and lift of BAE have an algebraic origin. From this viewpoint, assumption \textbf{A2} seems unnecessary. Indeed, all of the above constructions can be appropriately modified to accommodate the boundary case corresponding to \textbf{A1} and
\begin{description}
\item[A2$'$] $k=1,r$.
\end{description}
The necessary condition~\eqref{eq:wt_ker} for a non-vanishing $Q$--cohomology of a $\gl(M|N)$ Bethe vector $\omega$ implies that reducible solution of $\gl(M|N)$ BAE must have $\nu^{(2)}=L$ if $k=1$ and $\nu^{(r-1)}=L$ if $k=r$.
The reduced BAE must have a form~\eqref{eq:BAE_short} corresponding to simple root systems $\Delta_0^{M-1|N-1}=\{\alpha_3,\dots,\alpha_r\}$ and $\Delta_0^{M-1|N-1}=\{\alpha_1,\dots,\alpha_{r-2}\}$ induced by evaluating the subquotient $\big(\ker_{\langle\alpha_k,-\rangle} \Delta_0^{M|N}\big)/\mathbb{C}\alpha_k$ for $k=1$ and $k=r$ respectively.  
To satisfy these requirements one must replace \textbf{R1--R2} with
\begin{description}
\item[R$'$] $\{x_j^{(2)}\}_{j=1}^L=\{y_a\}_{a=1}^L$ for  $k=1$ and 
$\{x_j^{(r-1)}\}_{j=1}^L=\{\bar{y}_a\}_{a=1}^L$ for  $k=r$.
\end{description}
Invoking again claim~\ref{claim:coh} and the arguments that follow it, we conclude that the lift of a $\gl(M-1|N-1)$ Bethe vector to a $\gl(M|N)$ Bethe vector must be unique and satisfies $\nu^{(1)}=L$ for $k=1$ and $\nu^{(r)}=L$ for $k=r$.
Finally, the corresponding lifted solution can be evaluated explicitly if $\alpha_2$ is odd for $k=1$ and  $\alpha_{r-1}$ is odd for $k=r$
\begin{equation*}
 \{x^{(1)}_j\}_{j=1}^L = \{x^{(3)}_j\}_{j=1}^L\ ,\qquad \{x^{(r)}_j\}_{j=1}^L = \{x^{(r-2)}_j\}_{j=1}^L\ .
\end{equation*}

Subsectors in integrable systems are not a new phenomenon. For example the homogeneous $\gl(2n+1)$ spin chain $(V\otimes V^*)^L$ of sec.~\ref{sec:formalism} contains a subsector corresponding to the $2L$--th tensor power of the fundamental representation of $\osp(1|2n)$, see \cite{Saleur:2001cw}.
We shall provide more examples in sec.~\ref{sec:BAE}.
However, all these examples  lack by far the generality of the restriction and lifting procedures we have described.
This is because in our case there is an algebraic mechanism behind which relates the representation theories of $\{\gl(n+N|N)\}_{N\in\mathbb{Z}^+}$ Lie superalgebras.
The same cohomological reduction mechanism exists for any Lie superalgebra except $\osp(1|2n)$.
Therefore there is no doubt that the same restriction and lift phenomena occur also for general $\osp(R|2S)$ integrable spin chains.
Moreover, the embedding of spectra~\eqref{eq:embed_spec} was already observed on a (non-integrable) $\osp(2S+2|2S)$ spin chain in \cite{Candu:2008vw}.

We come back now to the spin chain of sec.~\ref{sec:formalism} and
answer the question about how excited w.r.t. each other are the subsectors~\eqref{eq:embed_spec}.

%%%%%%%%%%%%%%%%%%%%%%%%%%%%%%%%%%%%%%%%%%%%%%
%%%%%%%%%%%%%%%%%%%%%%%%%%%%%%%%%%%%%%%%%%%%%%

%\section{Bethe vectors}
%\label{sec:bv}
%%%%%%%%%%%%%%%%%%%%%%%%%%%%%%%%%%%%%%%%%%%%%%
%%%%%%%%%%%%%%%%%%%%%%%%%%%%%%%%%%%%%%%%%%%%%%

%%%%%%%%%%%%%%%%%%%%%%%%%%%%%%%%%%
%%%%%%%%%%%%%%%%%%%%%%%%%%%%%%%%%%
%%%%%%%%%%%%%%%%%%%%%%%%%%%%%%%%%%

\section{Vacuum and low lying excitations}
\label{sec:BAE}

%%%%%%%%%%%%%%%%%%%%%%%%%%%%%%%%%%
%%%%%%%%%%%%%%%%%%%%%%%%%%%%%%%%%%
%%%%%%%%%%%%%%%%%%%%%%%%%%%%%%%%%%

In this section we present numerical evidence showing that all integrable $\gl(n+N|N)$ spin chains~\eqref{eq:ham} with $n$ fixed have the same vacuum energy.
In view of the embedding of spectra~\eqref{eq:embed_spec}, we introduce the notion of degree of an excitation, which is the smallest value of $N$ for which it appears in $\spec H_{n+N|N}$.
We then proceed to classify the excitations of degree 0 and 1,
and present the
%%%%%
form of the
%%%%% 
numerical solutions of BAE reproducing them.

%%%%%%%%%%%%%%%%%%%%%%%%%%%%%%%%%%%%%%%%%%%
\subsection{Bosonic lift}
\label{sec:bos_lift}
%%%%%%%%%%%%%%%%%%%%%%%%%%%%%%%%%%%%%%%%%%%

From~\eqref{eq:embed_spec}, the vacuum energy of the spin chain is greater or equal to the vacuum energy of the algebraic Hamiltonian.
Supposing they coincide, from tables~(\ref{tab:1}, \ref{tab:2}) and the general discussion of sec.~\ref{sec:num_diag}
it follows that the vacuum state in the spin chain is a $\gl(n+N|N)$ invariant tensor.
According to eq.~\eqref{eq:weight_BV1}, the number of roots in the vacuum state should then be $\nu^{(k)}=L$.
Numerically, we confirm %in tab.~\ref{tab:3}
that there is indeed a solution to the $\gl(n+N|N)$ BAE~\eqref{eq:BAE_short} with $\nu^{(k)}=L$ which reproduces the vacuum energy of the algebraic Hamiltonian $\mathsf{H}\in B_{L,L}(n)$.
According to the terminology of sec.~\ref{sec:rl}, this is the $\gl(n+N|N)$ lift of the vacuum solution for the purely ``bosonic'' $\gl(n)$ chain.
The latter is characterized by $n-1$ seas of real roots without holes.

The low rank cases $\gl(2|1)$ and $\gl(3|1)$ have to be treated independently.
For the $\gl(2|1)$ case it was realized in \cite{Essler:2005ag} that the vacuum energy of the Hamiltonian~\eqref{eq:ham} is exactly $-4L$ and the corresponding vacuum solution in the grading $\Sigma=\{+,-,+\}$ is highly degenerated $x^{(1)}_j=x^{(2)}_j=0$.
This solution obviously does not make sense in the framework of the ABA. 
The problem persists in all other gradings as well.
A regularized vacuum Bethe vector can be defined by introducing arbitrary inhomogeneities for every $V$ and $V^*$ site in the chain.
Then, according to eqs.~\eqref{eq:freezing2}, the vacuum solution is entirely fixed by these inhomogeneities and the vacuum Bethe vector can be constructed.
The vacuum state of the homogeneous chain is then defined by a limiting procedure.
The $\gl(3|1)$ spin chain has the same problem, again in all the gradings, and the same cure.

We now realize  the embedding $\spec H_{n|0}\subset \spec H_{n+N|N}$ explicitly.
Following~\cite{Candu:2010yg}, choose an odd element $Q\in\gl(n+N|N)$ that squares to zero, has rank $N$ and defines the $\gl(n)$ spin chain $(V\otimes V^*)^{\otimes L}$ as the $Q$-cohomology of the $\gl(n+N|N)$ spin chain $(V\otimes V^*)^{\otimes L}$. 
For this choice of $Q$, one can perform the lift of all $\gl(n)$ Bethe vectors to $\gl(n+N|N)$ Bethe vectors as described in sec.~\ref{sec:rl}.
As we shall see in a moment, the lift depends on the choice of $Q$.
We call this part of the spectrum the \emph{bosonic lift}. The latter is straightforward to understand in terms of $\gl(n)$ spin chain excitations \cite{Andrei:1979un, Andrei:1979wy, Andrei:1979sq, Doikou:1998xi}.

\begin{figure}[t]%
\psfrag{l}{$L$}
\psfrag{d}{$\cdots$}
\psfrag{a}{$P$}
\psfrag{b}{$R$}
\psfrag{q1}{$Q_1$}
\psfrag{q2}{$Q_2$}
\psfrag{a1}{$\alpha_m$}
\psfrag{a2}{$\alpha_{m+2N+1}$}
\centerline{\includegraphics[scale=1.0]{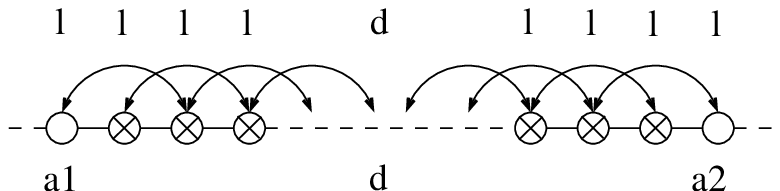}\hspace{0.5cm}
\includegraphics[scale=1.0]{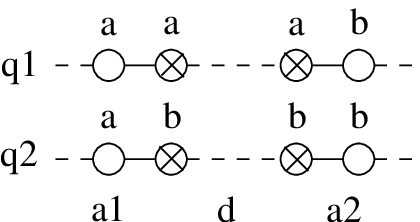}}
\caption{The grading $[+^{m}(+-)^{N}+^{n-m}]$, $1<m<n-1$ is such that there is a chain of $2N$ odd roots. On the left: lift of the $\gl(n)$ vacuum to the $\gl(n+N|N)$ vacuum. On the right: Bethe roots for Bethe vectors in the bosonic lift w.r.t. the two choices~\eqref{eq:two_q_choices} of $Q$.}%
\label{fig:bos_lift}%
\end{figure}

How to characterize the $\gl(n+N|N)$ symmetry of the excitations that belong to the bosonic lift?
We say  that the component $w_i$ of a $\gl(n+N|N)$ weight $w=\sum_{i=1}^{M+N}w_i\epsilon_i$ is bosonic if $|i|\equiv 0$ and fermionic if $|i|\equiv 1$. 
According to sec.~\ref{sec:rl}, if a $\gl(n+N|N)$ Bethe vector $\omega$ is the lift of a $\gl(n)$ Bethe vector, that is $\omega$ has degree 0, then it should be possible (in some grading) to write its weight in the form $\wt(\omega)=\sum_{i=1}^{M+N}w_i\epsilon_i$ with no fermionic components and at most $n$ bosonic components.
%Let $k$($l$) denote the number of positive (negative) components $w_i$.
Equivalently, the weight of a Bethe vector $\omega$ of degree 0 must be representable (in some grading) by highest weight Young supertableaux of $\gl(n)$--admissible shape $(\lambda,\mu)$
\begin{equation}\label{eq:adm_shape_bos}
 \lambda'_1+\mu'_1\leq n\ .
\end{equation}
These shapes are $\Sigma$--admissible and most obviously $\gl(n)$ reducible w.r.t. the gradings
\begin{equation}\label{eq:some_special_grad}
\Sigma=\{\overbrace{+,\dots,+}^m,\overbrace{+,-,\dots,+,-}^{2N},\overbrace{+,\dots,+}^{n-m}\}=[+^m(+-)^N+^{(n-m)}]\ ,
\end{equation}
where $\lambda'_1\leq m\leq n-\mu'_1$.
There are many choices for the nilpotent element to define the restriction and lift in the grading~\eqref{eq:some_special_grad}.
For instance, the two possibilities
\begin{align}\label{eq:two_q_choices}
 Q_1 \in \mathbb{C}E_{m+2,m+1}\oplus \mathbb{C} E_{m+4,m+3}\oplus\cdots \oplus \mathbb{C}E_{m+2N,m+2N-1}\\
Q_2 \in \mathbb{C}E_{m+3,m+2}\oplus \mathbb{C} E_{m+5,m+4}\oplus\cdots \oplus \mathbb{C}E_{m+2N+1,m+2N}\notag
\end{align}
are represented in fig.~\ref{fig:bos_lift}. Notice that for a fixed $Q$ the lift is unique, although it is different for different $Q$'s. 
Multiple gradings of the type~\eqref{eq:some_special_grad} must be considered in order to recover the full bosonic lift.

To conclude, we write down explicitly the vacuum solution in the grading $[+^{m} (+-)^{N} +^{n-m} ]$ with $1<m<n-1$
\begin{align}\label{eq:vac_sol}
 x^{(m)}_j&=x^{(m+2)}_j=\dots=x^{(m+2N)}_j\\ \notag
x^{(m+1)}_j&=x^{(m+3)}_j=\dots=x^{(m+2N+1)}_j\ ,\qquad j=1,\dots,L\ ,
\end{align}
where the roots $\{x^{(l)}_j\}_{j=1}^L$, $l=1,\dots,m,m+2N+1,\dots,n+2N-1$ correspond to the $\gl(n)$ vacuum solution. We represent the lift of the bosonic vacuum  in fig.~\ref{fig:bos_lift}.
We have already mentioned, see eq.~\eqref{eq:freezing2}, how to treat the special cases $\gl(2|1)$ and $\gl(3|1)$ by introducing arbitrary inhomogeneities.

%%%%%%%%%%%%%%%%%%%%%%%%%%%%%%%%%%%%%%%%%%%
\subsection{Degree one excitations}
\label{sec:deg1}
%%%%%%%%%%%%%%%%%%%%%%%%%%%%%%%%%%%%%%%%%%%

\subsubsection{Classification}

\begin{figure}[t]%
\psfrag{d}{$\cdots$}
\psfrag{a}{$P$}
\psfrag{b}{$Q$}
\psfrag{c}{$R$}
\psfrag{e}{$S$}
\psfrag{a1}{$\alpha_1$}
\psfrag{a2}{$\alpha_{m-n}$}
\psfrag{a3}{$\alpha_{m-n+2N}$}
\psfrag{a4}{$\alpha_{n+2N-1}$}
\psfrag{l1}{$L-1$}
\psfrag{l2}{$L-m+1$}
\psfrag{l3}{$L-m$}
\psfrag{l4}{$L-m-1$}
\psfrag{l5}{$L-m$}
\psfrag{l6}{$L-m+1$}
\psfrag{l7}{$L-1$}
\psfrag{g1}{$\gl(2m|1)$}
\psfrag{g2}{$\gl(2m+1|1)$}
\psfrag{b1}{$\alpha_m$}
\psfrag{b2}{$\alpha_{m+2N}$}
\centerline{\includegraphics[scale=1.0]{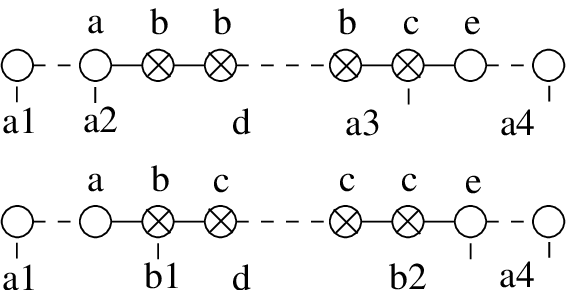}\hspace{1.5cm}
\includegraphics[scale=1]{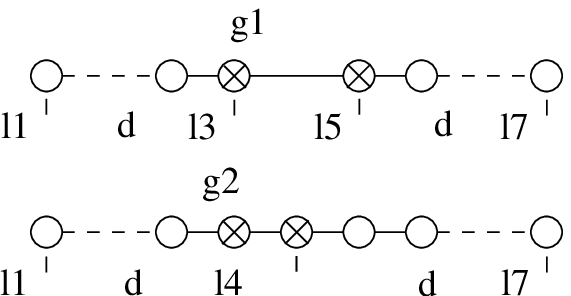}
}
\caption{Root numbers for $\gl(n+N|N)$ excitations of degree 1 and shape $(\lambda,\mu)$ satisfying, top left, eq.~\eqref{eq:deg_1a} w.r.t. some grading $[(+)^{n-m}(+-)^{2N}(+)^m]$ with $\mu_1'\leq m\leq n+1-\lambda_2'$ or, bottom left, eq.~\eqref{eq:deg_1b} w.r.t. some grading $[(+)^m(-+)^{2N}(+)^{n-m}]$ with  $\lambda_1'\leq m\leq n+1-\mu_2'$. 
On the right: root numbers for the lowest lying degree 1 excitation.}%
\label{fig:deg1}%
\end{figure}

Fix an odd element $Q\in\gl(n+N|N)$ that squares to zero, has rank $N-1$ and defines the $\gl(n+1|1)$ spin chain $(V\otimes V^*)^{\otimes L}$ as the $Q$-cohomology of the $\gl(n+N|N)$ spin chain $(V\otimes V^*)^{\otimes L}$. 
Excitations of degree 1 correspond to $\gl(n+N|N)$ Bethe vectors which: i) are non-reducible to $\gl(n)$ Bethe vectors and ii) are lifted $\gl(n+1|1)$ Bethe vectors.
%~\footnote{
%Our degree of an excitation is related to the \emph{degree of atypicality} of~\cite{serganova}.
%If $\omega$ is a $\gl(n+N|N)$ Bethe vector, then $N-\deg(\omega)$ is the degree of atypicality of $\wt(\omega)$.}
We shall call the corresponding Bethe vectors also of degree 1.
According to sec.~\ref{sec:rl}, the weight of a Bethe vector $\omega$ of degree 1 admits a representation (in some grading) of the form $\wt(\omega)=\sum_{i=1}^{M+N}w_i \epsilon_i$ with at most $n+1$ bosonic components and at most one fermionic component.
Notice that due to the claim~\ref{claim:coh} of sec.~\ref{sec:rl}, there are also excitations of degree 1 with the same weight as excitations of degree 0, that is no fermionic components and at most $n$ bosonic components.
Taking into account sec.~\ref{sec:tens} as well, the weight of Bethe vectors of degree 1 must be representable (in some grading) by highest weight Young supertableaux of $\gl(n+1|1)$--admissible shapes $(\lambda,\mu)$,
that is at least one of the two conditions holds
\begin{align}\label{eq:deg_1a}
 \lambda'_1+\mu'_2&\leq n+1\\
\lambda'_2+\mu'_1&\leq n+1\ .
\label{eq:deg_1b}  
\end{align}
Shapes~\eqref{eq:deg_1a} are $\Sigma$--admissible and most obviously $\gl(n+1|1)$ reducible w.r.t. the gradings $[(+)^{n-m}(+-)^{2N}(+)^m]$, such that $\mu_1'\leq m\leq n+1-\lambda_2'$, and the choice  $Q\in \mathbb{C}E_{m-n+3,m-n+2}\oplus\dots \oplus \mathbb{C}E_{m-n+2N-1,m-n+2N-2}$.
Shapes~\eqref{eq:deg_1b} are $\Sigma$--admissible and most obviously $\gl(n+1|1)$ reducible w.r.t. the gradings $[(+)^m(-+)^{2N}(+)^{n-m}]$, such that $\lambda_1'\leq m\leq n+1-\mu_2'$, and the choice  $Q\in\mathbb{C}E_{m+3,m+2}\oplus\cdots \oplus E_{m+2N-1,m+2N-2}$.
The corresponding root configurations are represented in fig.~\ref{fig:deg1} on the left.

\subsubsection{Low lying excitations}
\label{sec:low_lying_deg1}

Assumption~2 of sec.~\ref{sec:spec} and tab.~\ref{tab:2} clearly indicates that the lowest lying excitation of the $B_{L,L}(n)$ algebraic Hamiltonian which is of degree higher then 0 lies in $\Delta_{L,L}(1^k,1^k)$ with $k=[\tfrac{n}{2}]+1$.
Equivalently, the lowest lying excitation of the $\gl(n+N|N)$ spin chain Hamiltonian which is of degree higher then 0 is a traceless tensor $t_0(1^k,1^k)$. Eqs.~(\ref{eq:deg_1a}, \ref{eq:deg_1b}) imply
this excitation is of degree 1.
To reproduce it from $\gl(n+1|1)$ BAE we have chosen the following gradings
\begin{equation}\label{eq:grad_special}
 \Sigma=\begin{cases}
[+^{m}-+^{m}]\phantom{^{+1}}\ ,& n=2m-1\\
[+^{m} - +^{m+1}]\ ,& n=2m
        \end{cases}
\end{equation}
The corresponding root numbers are represented in fig.~\ref{fig:deg1} on the right.
%We present the numerical solutions of BAE in tab.~\ref{tab:4}.
%%%%%%%%%%%%%%%%%%%%%%%%%%%%%%%%%%%%%%%%%
We have performed extensive numerical calculations on the form of the solution of Bethe ansatz equations reproducing the lowest lying excitations of degree 1.~\footnote{We will show a plot of the root distributions in the complex place for low lying excitations of degree 1 in a future version of this eprint.}
%%%%%%%%%%%%%%%%%%%%%%%%%%%%%%%%%%%%%%%%%
%We noticed that,
Compared to usual Lie algebra spin chains, this excitation looks very strange, to say the least.
Strikingly, for $n$ odd the roots $\{x_j^{(m)}\}_{j=1}^{L-m}$, $\{x_j^{(m+1)}\}_{j=1}^{L-m}$ are always complex, while for $n$ even the roots
$\{x^{(m+1)}_j\}_{j=1}^{L-m-1}$ are always complex.
We recall that all roots are real for the  vacuum solution~\eqref{eq:vac_sol}.
Thus, this excitation is not constructed as usual by making ``minor'' modifications to the vacuum solution.

Another strange feature is the non-selfconjugacy of the solutions, at least in general.
This is a consequence of the non-hermiticity of the Hamiltonian~\eqref{eq:ham}.~\footnote{See~\cite{vlad} for a proof of selfconjugacy in the case of hermitian Lie algebra spin chain Hamiltonians.}
Such solutions have been investigated already in the study~\cite{Essler:2005ag} of the $\gl(2|1)$ spin chain, but also encountered in different contexts dealing with non-hermitian Hamiltonians~\cite{Saleur:2000bq}. 
Instead, all lowest lying degree 1 solutions
%%%%%%%%%%%%%%%%%
%in tab.~\ref{tab:4}
%%%%%%%%%%%%%%%%%%%%
are invariant w.r.t. a modified conjugation symmetry
\begin{align}\label{eq:conj_1}
 \gl(2m|1):&& \{x^{(k)*}_j\}_{j=1}^{\nu^{(k)}}&=\{x^{(n+1-k)}_j\}_{j=1}^{\nu^{(n+1-k)}}\\
 \gl(2m+1|1):&& \{x^{(k)*}_j\}_{j=1}^{\nu^{(k)}}&=\{x^{(n+1-k)}_j\}_{j=1}^{\nu^{(n+1-k)}}\ ,\quad k\neq m+1\ .\label{eq:conj_2}
\end{align}
The symmetry~\eqref{eq:conj_2} looks surprising, because the Dynkin diagram in fig.~\ref{fig:deg1} is not symmetric w.r.t. the transposition $\alpha_{k}\mapsto\alpha_{n+1-k}$. However, notice that the roots $\{x_j^{(k)}\}_{j=1}^{\nu^{(k)}}$, $k\neq m+1$ together with the particle hole transformed roots $\{x_j^{(m+1)}\}_{j=1}^{\nu^{(m+1)}}$ also solve the BAE corresponding to the transposed Dynkin diagram of fig.~\ref{fig:deg1}, see~\cite{Tsuboi:1998ne} for details. 
For $\nu^{(m+1)}$ even, the solutions
%%%%%%%%%%%%%%%%%%%%%%%
%in tab.~\ref{tab:4}
%%%%%%%%%%%%%%%%%%%%%%%
are (exceptionally) selfconjugate
\begin{equation}\label{eq:exceptional}
 \nu^{(m+1)} \quad \text{even}:\qquad \{x^{(k)*}_j\}_{j=1}^{\nu^{(k)}}=\{x^{(k)}_j\}_{j=1}^{\nu^{(k)}}\ , \quad k=1,\dots, n+1\ .
\end{equation}

The following picture of the distribution of roots in the complex plane in the thermodynamic limit is emerging from
our numerical analysis
%%%%%%%%%%%%%%%%%%%%
%tab.~\ref{tab:4}
%%%%%%%%%%%%%%%%%%%%%
%
\begin{align}\label{eq:ideal_sol_1}
\gl(2m|1) &:& &\begin{cases}\begin{matrix}
 x^{(k)}_j &=& x^{(2m+1-k)}_j&,& x^{(k)*}_j = x^{(k)}_j&,& k \neq m,m+1  \\[5pt]
x^{(m)}_j &=& x^{(m+1)*}_j &=& x^{(\pm)}_j \pm \frac{i}{4} & , &  x_j^{(\pm)*}=x_j^{(\pm)}
\end{matrix}\end{cases}\\[5pt]
\gl(2m+1|1) &:& &\begin{cases} \begin{matrix}
 x^{(k)}_j &=& x^{(2m+2-k)}_j&, &  x^{(k)*}_j = x^{(k)}_j & , & k \neq m+1\\[5pt]
x^{(m+1)}_j &=&x^{(\pm)}_j \pm \frac{i}{2} & ,& x_j^{(\pm)*}=x_j^{(\pm)} & &  
\end{matrix}
\end{cases}\ . \label{eq:ideal_sol_2}
\end{align}
The $\gl(2m|1)$ complex root configurations where called $\pm$-strange strings in~\cite{Essler:2005ag}.

We see that the lowest degree 1 excitation looks pretty complicated.
What about other degree 1 excitations? We have verified that properties~(\ref{eq:conj_1}--\ref{eq:conj_2}) hold for all low lying excitations with the same root numbers as in fig.~\ref{fig:deg1} on the right and, more then that, for all low lying solutions with symmetric configurations of the root numbers $\nu^{(k)}=\nu^{(n+1-k)}$.
Many of these excitations seem to tend to the form~(\ref{eq:ideal_sol_1}, \ref{eq:ideal_sol_2}) in the thermodynamic limit.
These are the best understood excitations for which we shall give a continuum description in the next section.
We have also noticed that among the low lying excitations with $\nu^{(m+1)}$ even, there is always a \emph{subset} of selfconjugate solutions of the type~\eqref{eq:exceptional}. These have interesting properties, as we shall see in a moment.

To complete the overview of low lying degree 1 excitations, let us mention that we have also identified solutions with symmetrical configurations of the 
root numbers, which do not seem to tend to the form~(\ref{eq:ideal_sol_1}, \ref{eq:ideal_sol_2}) in the thermodynamic limit.
For these solutions we have nothing to say, although their study might prove crucial in constructing an $S$-matrix description of the continuum theory in large volume.
The situation is even worse for solutions with non symmetrical configurations of the root numbers, because $\pm$-strange string configurations cannot be clearly defined. 
%We present the solutions of BAE in the grading~\eqref{eq:grad_special}, reproducing  other low lying excitations of degree 1 in tab.~\ref{tab:5}.

\subsubsection{Selfconjugate solutions}

As we have said, for $\nu^{(m+1)}$ even and  $\nu^{(k)}=\nu^{(n+1-k)}$, there is always a subset of selfconjugate solutions of the type~\eqref{eq:exceptional}.
It is straightforward to show that the $\gl(2m|1)$ BAE~\eqref{eq:BAE_short} subject to the constraints~(\ref{eq:conj_1}, \ref{eq:exceptional}) are equivalent to the BAE of the $\so(2m+1)$ fundamental spin chain of length $L$ with periodic boundary conditions for the first $m-1$ types of roots and antiperiodic boundary conditions for the spinorial roots.
This is the generalization of the Takhtajan-Babujian subsector of the $\gl(2|1)$ spin chain of~\cite{Essler:2005ag}.
The eigenvalues of the integrable Hamiltonian for this $\so(2m+1)$ spin chain are given by eq.~\eqref{eq:spec_ham_mom} subject to the constraint~\eqref{eq:exceptional}.
While this correspondence between solutions of $\gl(2m|1)$ and $\so(2m+1)$ BAE is quite curious, the remarkable thing is that it extends to a correspondence between weakly excited solutions on both sides.
In particular, the vacuum solution of the $\so(2m+1)$ fundamental chain  in the thermodynamic limit~\cite{Martins:1990ca} (plus holes) is compatible with the form~\eqref{eq:ideal_sol_1} when the constraint~\eqref{eq:exceptional} is taken into account.
We shall see how this correspondence can be used effectively in the next section.

The situation is quite different for $\gl(2m+1|1)$. Notice that assumptions \textbf{A1--A2} and \textbf{R1} of sec.~\ref{sec:rl} hold for solutions of $\gl(2m+1|1)$ BAE~\eqref{eq:BAE_short} satisfying~(\ref{eq:conj_2}, \ref{eq:exceptional}) and, therefore, these can be restricted to solutions of $\gl(2m)$ BAE.
However, the latter are very special,  because in addition to satisfying the $\gl(2m)$ BAE they must also admit multiple $\gl(2m+1|1)$ lifts.
According to the claim 2 of sec.~\ref{sec:rl}, $\gl(2m+1|1)$ lifted \emph{Bethe vectors} correspond exclusively to $\gl(2m)$ solutions that do not admit multiple lifts.
Presumably, what happens is that the additional conditions satisfied by these $\gl(2m)$ solutions obtained by restriction ensure the vanishing of the corresponding $\gl(2m)$ Bethe vectors.
We have checked that
%%%%%%%%%%%%%%%%%%%%%%%%%%%%
%besides the solutions presented in tab.~\ref{tab:4},
%%%%%%%%%%%%%%%%%%%%%%%%%%%%%%%%%
there are many other low lying solutions satisfying the constraints~(\ref{eq:conj_2}, \ref{eq:exceptional}).
However, it is most confusing that many low lying solutions that \emph{do not} satisfy the constraints~(\ref{eq:conj_2}, \ref{eq:exceptional}) in finite volume seem to have  a thermodynamic limit~\eqref{eq:ideal_sol_2} that does satisfy the constraints~(\ref{eq:conj_2}, \ref{eq:exceptional}).
We are not sure how to interpret this behavior,
although we are tempted to believe this indicates that the large volume limit is a subtle issue.
Therefore, we concentrate in the next section solely on the continuum description of $\gl(2m|1)$ degree 1 excitations of the type~(\ref{eq:conj_1}, \ref{eq:ideal_sol_1}).

%%%%%%%%%%%%%%%%%%%%%%%%%%%%%%%%%%%%%%%%%%%
%%%%%%%%%%%%%%%%%%%%%%%%%%%%%%%%%%%%%%%%%%%

\section{Continuous limit}
\label{sec:cl}
%%%%%%%%%%%%%%%%%%%%%%%%%%%%%%%%%%%%%%%%%%%
%%%%%%%%%%%%%%%%%%%%%%%%%%%%%%%%%%%%%%%%%%%

In this section we shall consider the spectrum of the spin chain of sec.~\ref{sec:formalism} in the thermodynamic limit $L\to \infty$.
The homogeneous spin chains are gapless and the only thermodynamic quantity, as far as the spectrum is concerned, is the vacuum energy per site. The latter does not provide any insight into the the continuous limit of the chain, which is expected to be governed by a conformal field theory (CFT). 
Probing this CFT requires computing scaling corrections to the spectrum~\cite{Cardy:1986ie}.
However, these are much harder to compute, e.g. see~\cite{pearce}, then thermodynamic quantities. Therefore, gaining insight into the CFT solely from the lattice is quite non-trivial.
To avoid such complications, one can introduce a smooth gap in the spin chain.
The continuous limit is then expected to be a massive deformation of the CFT.
Many interesting quantities, such as the $\beta$-function, the particle spectrum, $S$-matrices, can now be computed in the thermodynamic limit.
The standard way to generate a gap in the homogeneous spin chains is by introducing an alternating inhomogeneity $\Lambda$ in the monodromies~(\ref{eq:mon_mat_a}, \ref{eq:mon_mat_ab}), called staggering~\cite{Faddeev:1996iy}. The source terms of the homogeneous BAE~\eqref{eq:BAE_short} then change to 
\begin{equation*}
 \left(\frac{x^{(k)}_j+i/2}{x^{(1)}_j-i/2}\right)^L\mapsto \left(\frac{x^{(k)}_j-\Lambda/2+i/2}{x^{(k)}_j-\Lambda/2 -i/2}\right)^{L/2} \left(\frac{x^{(k)}_j+\Lambda/2+i/2}{x^{(k)}_j+\Lambda/2 -i/2}\right)^{L/2}\ ,\qquad k=1,r\ ,
\end{equation*}
while the eigenvalues of the new energy and momentum operators become
\begin{align}\label{eq:spec_ham_mom_stag}
 E &= -\sum_{j=1}^{\nu^{(1)}}\frac{\frac{\sigma_1}{2}}{\big(x^{(1)}_j -\frac{\Lambda}{2}\big)^2+\frac{1}{4}}+\frac{\frac{\sigma_1}{2}}{\big(x^{(1)}_j -\frac{\Lambda}{2}\big)^2+\frac{1}{4}}
-\sum_{j=1}^{\nu^{(r)}}\frac{\frac{\sigma_{M+N}}{2}}{\big(x^{(r)}_j-\frac{\Lambda}{2}\big)^2+\frac{1}{4}}+\frac{\frac{\sigma_{M+N}}{2}}{\big(x^{(r)}_j+\frac{\Lambda}{2}\big)^2+\frac{1}{4}}\\
P &\equiv  \frac{\sigma_1}{2}\sum_{i=1}^{\nu^{(1)}}\theta_1\left(x^{(1)}_j-\frac{\Lambda}{2}\right)+\theta_1\left(x^{(1)}_j+\frac{\Lambda}{2}\right)+ \frac{\sigma_{M+N}}{2}\sum_{i=1}^{\nu^{(r)}}\theta_1\left(x^{(r)}_j-\frac{\Lambda}{2}\right)+\theta_1\left(x^{(r)}_j+\frac{\Lambda}{2}\right)\ , \notag
\end{align}
where $\sigma_1=(-1)^{|1|}$, $\sigma_{M+N}=(-1)^{|M+N|}$ and we have assumed $L$ to be even. These expressions reduce the the previous ones~\ref{eq:spec_ham_mom} in the limit $\Lambda\to0$.

In this section we define and study the continuous limit of the staggered $\gl(n+N|N)$ chains of sec.~\ref{sec:formalism}. First we show that the bosonic lift of the staggered chain is described in the continuous limit by the $\gl(n)$ Gross-Neveu (GN) model~\cite{Andrei:1979un, Andrei:1979wy, Andrei:1979sq}.
This strongly suggests that the continuous limit of the staggered spin chain coincides with the $\gl(n+N|N)$ GN model, because the latter contains the $\gl(n)$ GN model as a cohomological subsector~\cite{Candu:2010yg}. The identification of the continuous limit is the main result of the paper.
In the second subsection we explore the particle content of the $\gl(2m|1)$ GN model which does not lie in the $\gl(2m-1)$ GN model lift, that is the bosonic lift.

%%%%%%%%%%%%%%%%%%%%%%%%%%%%%%%%%%%%%%%%%%%
\subsection{Bosonic lift}
%%%%%%%%%%%%%%%%%%%%%%%%%%%%%%%%%%%%%%%%%%%

As we have explained in sec.~\ref{sec:rl} and~\ref{sec:bos_lift}, the solutions of $\gl(n)$ BAE corresponding to non-vanishing Bethe vectors admit a unique lift to solutions of $\gl(n+N|N)$ BAE and Bethe vectors. We called this subsector of the $\gl(n+N|N)$ spin chain the bosonic lift.
The uniqueness of the lift implies that the fermionic roots of the lifted solutions of $\gl(n+N|N)$ BAE, in the form schematically represented in fig.~\ref{fig:bos_lift}, are uniquely determined by the even roots.
By the restriction procedure of sec.~\ref{sec:rl}, the latter solve $\gl(n)$ BAE.
Therefore, all dynamical degrees of freedom (associated to holes) are determined by the even roots.
The odd roots behave as auxiliary quantities useful for defining lifted Bethe vectors.
So, we conclude that the BAE of the $\gl(n)$ spin chain describe \emph{entirely} the bosonic lift of the $\gl(n+N|N)$ spin chain.
We recall some of the old results on the continuum limit of  $\gl(n)$ spin chains and then reinterpret them in the $\gl(n+N|N)$ context.

\subsubsection{Vacuum energy}\label{sec:vac_en}

\begin{figure}[t]%
\psfrag{1}{$1$}
\psfrag{2}{$2m$}
\psfrag{3}{$2m-2$}
\psfrag{4}{$m$}
\psfrag{5}{$m-1$}
\psfrag{d}{$\cdots$}
\psfrag{g1}{$\gl(2m|1)\, (V\otimes V^*)^{\otimes L}$}
\psfrag{g2}{$\gl(2m-1)\, (V\otimes V^*)^{\otimes L}$}
\psfrag{g3}{$\so(2m+1)\, V^{\otimes L}$}
\psfrag{g4}{$\osp(1|2m-2)\, V^{\otimes L}$}
\centerline{\includegraphics[scale=1]{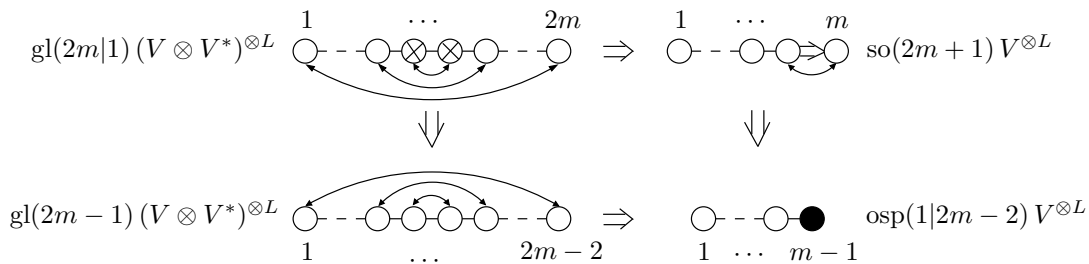}\hfill}
\caption{Dynkin diagrams represent BAE for the corresponding chains.
Double arrows indicate the equality of the corresponding set of roots.
}%
\label{fig:emb}%
\end{figure}

According to sec.~\ref{sec:bos_lift}, the vacuum energy of the $\gl(n+N|N)$ spin chain coincides exactly with the vacuum energy of the $\gl(n)$ spin chain.
The vacuum energy $e_{\infty}=\lim_{L\to \infty} \tfrac{E_{\text{vac}}}{2L}$ of the homogeneous $\gl(n)$ spin chain $V^{\otimes L}$ was computed in~\cite{Sutherland:1975vr}. This calculation is easily generalized for the staggered $\gl(n)$ spin chains $(V\otimes V^*)^{\otimes L}$
\begin{align}\notag
e_{\infty}&= -\frac{1}{2n}\left[\psi(1+\tfrac{i\Lambda}{n})-\psi(\tfrac{1}{n}+\tfrac{i\Lambda}{n})+\psi(\tfrac{1}{2}+\tfrac{1}{n}+\tfrac{i\Lambda}{n})-\psi(\tfrac{1}{2}+\tfrac{i\Lambda}{n}) +c.c\right] \\
&\phantom{=}-\frac{1}{n}\left[
\psi(1)-\psi(\tfrac{1}{n})+\psi(\tfrac{1}{2}+\tfrac{1}{n})-\psi(\tfrac{1}{2})\right] \ , \label{eq:vac_en}
\end{align}
where $\psi$ is the digamma function.
Inserting $n=1$ and $\Lambda=0$ we get the result $e_{\infty}=-4$ of~\cite{Essler:2005ag}.
For $n=2m-1$ the vacuum energy coincides exactly with the vacuum energy of the staggered $\so(2m+1)$ and $\osp(1|2m-2)$ fundamental chains of length $L$.
These coincidences where probably first noticed in~\cite{Saleur:2001cw, Martins:1995bb}.  The correspondence between the spectra of $\gl(2m|1)$, $\gl(2m-1)$, $\so(2m+1)$ and $\osp(1|2m-2)$ chains is easily understood at the level of BAE from fig.~\ref{fig:emb}.
The non-trivial feature of the correspondence is that it relates low lying with low lying excitations.
It is exactly this type of correspondences that lie at the heart of the dualities observed in~\cite{Saleur:2009bf}.

%%%%%%%%%%%%%%%%%%%%%%%%%%%%%%%%%%%%%%%
\subsubsection{Particle spectrum}
%%%%%%%%%%%%%%%%%%%%%%%%%%%%%%%%%%%%%%

Let us first recall some of the results on the continuous limit of $\gl(n)$ chains with real vacuum solution.
Going on the lines of~\cite{Andrei:1979wy}, the $\gl(n)$ staggered spin chain $(V\otimes V^*)^{\otimes L}$ can be shown to posses $n-1$ gapped excitations branches associated with the holes in the distributions of $n-1$ types of roots. For $\Lambda$ big, the gap is of order $e^{-\pi \Lambda/n}$. The continuous limit is performed by introducing a physical lattice spacing $a$ and then by taking the limit $a\to0, \Lambda\to\infty$ such that
\begin{equation}\label{eq:cont_limit}
m= a^{-1} e^{-\pi \Lambda/n} 
\end{equation}
remains fixed, defining the mass scale of the theory.
If we correct for the Fermi velocity $c=4\pi/n$, then in 
the continuous limit each excitation branch yields a relativistic dispersion relation with mass
\begin{equation}\label{eq:bos_particles}
 m_k = 2m\sin \frac{\pi k}{n},\qquad k=1,\dots,n-1\ .
\end{equation}
Therefore, we interpret it as a particle in some relativistic quantum field theory.
The lowest lying excitation of the spin chain corresponding to a traceless tensor $t_0(\lambda,\mu)$ of admissible shape~\eqref{eq:adm_shape_bos} is reproduced by a real solution of BAE with root and hole numbers fixed by the highest weight of the tensor~\cite{Doikou:1998xi}.
This excitation is associated to a scattering eigenstate in the field theory of energy
\begin{equation}\label{eq:lowest_lambla_mu}
 E_{\lambda,\mu}-E_{\text{vac}}=\sum_{i=1}^{\lambda_1} m_{\lambda'_i}\cosh \theta_i + \sum_{i=1}^{\mu_1} m_{\mu'_i}\cosh \bar{\theta}_i\ ,
\end{equation}
where $\theta$, $\bar{\theta}$ are the physical rapidities parametrizing the hole positions. 
Eq.~\eqref{eq:lowest_lambla_mu} suggests that the $k$-th particle~\eqref{eq:bos_particles} corresponds to either a covariant $1^k$ or contravariant $\bar{1}^k$ antisymmetric tensor.
This is confirmed by an $S$-matrix calculation~\cite{Andrei:1979sq, Doikou:1998xi}.
%, where now complex solutions play a crucial role.
The end result is the  $S$-matrix of the $\gl(n)$ Gross-Neveu model~\cite{PhysRevD.20.897}.
For instance, $1$ and $\bar{1}$ particles scatter with themselves with
the $S$-matrix
\begin{equation}\label{eq:smat_11}
 S_{11}(\theta)=S_{\bar{1}\bar{1}}(\theta)=Z_{11}(\theta)\frac{P + i n \theta/2\pi}{1 + i n \theta/2\pi}\ ,\qquad S_{1\bar{1}}=Z_{1\bar{1}}(\theta)\left(1-\frac{Q}{n/2+in\theta/2\pi}\right)\ ,
\end{equation}
where $P$ and $Q$ are permutation and Temperley-Lieb operators defined in sec.~\ref{sec:formalism} and 
\begin{equation*}
 Z_{11}(\theta)=\frac{\Gamma(1-1/n-i\theta/2\pi)\Gamma(1+i\theta/2\pi)}{\Gamma(1-1/n+i\theta/2\pi)\Gamma(1-i\theta/2\pi)}\ ,\ Z_{1\bar{1}}(\theta)=\frac{\Gamma(1/2-i\theta/2\pi)\Gamma(1/2-1/n+i\theta/2\pi)}{\Gamma(1/2+i\theta/2\pi)\Gamma(1/2-1/n-i\theta/2\pi)} 
\end{equation*}
are, so called, dressing factors.
The GN model coupling is related to the staggering $g=\Lambda^{-1}$ and eq.~\eqref{eq:cont_limit} yields its $\beta$-function. 

There is one important remark. Multiparticle states have zero total $\un(1)=\gl(n)/\sgl(n)$ charge, because the  spin chain $(V\otimes V^*)^{\otimes L}$ has zero $\un(1)$ charge.
Therefore, it is not possible to fix the $\un(1)$ charge of the individual particles in the spin chain. The allowable multiparticle configurations in the spin chain make the value of this charge irrelevant. However, the field theories with $\un(1)$ charged and uncharged particles~\eqref{eq:bos_particles} are clearly different: the first one has $2n-2$ irreducible multiplets of particles, while the latter, which is the GN model, has $n-1$ irreducible multiplets.

We now reinterpret the results of the above calculations in the framework of the bosonic lift of the $\gl(n+N|N)$ spin chain.
The continuous limit of the $\gl(n+N|N)$ spin chain is defined by the same eq.~\eqref{eq:cont_limit}. It already implies that the $\beta$-function of the continuous theory does not depend on $N$.
The $n-1$ excitation branches of the bosonic lift are interpreted again as relativistic particles of %the $\gl(n+N|N)$ GN model of 
mass~\eqref{eq:bos_particles}.
Eq.~\eqref{eq:lowest_lambla_mu} gives now the energy of a scattering eigenstate corresponding to a $\gl(n+N|N)$ traceless tensor $t_0(\lambda,\mu)$ of $\gl(n)$ admissible shape~\eqref{eq:adm_shape_bos}.
However now, it suggests that the bosonic lift contains $2n-2$ distinct multiplets of particles corresponding to the $\gl(n+N|N)$ covariant antisymmetric tensors $1,1^2,\dots,1^{n-1}$ and  their antiparticles, that is the contravariant antisymmetric tensors $\bar{1},\bar{1}^2,\dots,\bar{1}^{n-1}$.
Notice that the number of particles does not depend on whether they carry or not a $\un(1)$ charge, which again cannot be fixed in the spin chain. This is because the $\sgl(n+N|N)$ antisymmetric tensors  $1^k$  and $1^{n-k}$  are  not isomorphic anymore. 

The suggested $\gl(n+N|N)$ symmetry of the particles in the bosonic lift must be confirmed by an $S$-matrix calculation. While we do not know the full $S$-matrix, one can compute its \emph{restriction} to the bosonic lift.
The calculation is formally the same as for the $\gl(n)$ spin chain, because the BAE are the same.
Thus, the eigenvalues of the $\gl(n+N|N)$ spin chain $S$-matrix restricted to the bosonic lift coincide with the eigenvalues of the $\gl(n)$ spin chain $S$-matrix.
More precisely, the eigenvalue of the $\gl(n+N|N)$ spin chain $S$-matrix on a scattering eigenstate  corresponding to a $\gl(n+N|N)$ traceless tensor $t_0(\lambda,\mu)$ of $\gl(n)$-admissible shape does not depend on $N$.
Due to the fact that the $Q$-cohomology of a tensor product of $\gl(n+N|N)$ representations is the tensor product of their $Q$-cohomologies~\cite{Candu:2010yg} all these eigenvalues are compatible with our earlier assumption that the particles of the bosonic lift are co- and contravariant antisymmetric tensors of $\gl(n)$-admissible shape.

Finally, the scattering of particles in the bosonic lift can be computed exactly when no scattering eigenstates outside the bosonic lift are generated. 
Thus, particles $1$ and $\bar{1}$ scatter with the same $S$-matrix~\eqref{eq:smat_11}, where this time $P$ and $Q$ act on tensor products of $\gl(n+N|N)$ representations.
Notice that we did not assume any crossing or unitarity to derive $S_{1\bar{1}}$ from $S_{11}$.
The question now is what field theory can reproduce these $S$-matrices? Given its formal similarity with the $\gl(n)$ GN $S$-matrix, the most obvious candidate is the $\gl(n+N|N)$ GN model.
There are deeper reasons to believe this.
First of all, the $Q$-cohomology of the $\gl(n+N|N)$ GN model is the $\gl(n)$ GN model. In particular, their $\beta$-functions coincide~\cite{Candu:2010yg}.
Secondly, there is no doubt that the perturbative $1/n$ calculations of the $\gl(n)$ GN $S$-matrix can be generalized. While we did not carry out an honest perturbative $S$-matrix calculation as in~\cite{PhysRevD.20.897}, the result should certainly be~\eqref{eq:smat_11}.
This is because the Feynman rules are  $\gl(n+N|N)$ invariant tensors and their algebra, generated by tensor multiplication and contraction, is a representation of the walled Brauer algebra $B_{L,L}(n)$ of sec.~\ref{sec:wb_sub} and~\ref{sec:tens}. Thus, as long as one computes an invariant tensor, such as the $S$-matrix, starting from other invariant tensors, such as the Feynman rules, the result cannot depend on $N$.
Thirdly, the detailed analysis of~\cite{Essler:2005ag, Saleur:2006tf} suggests that the continuum limit of the homogeneous $\gl(2|1)$ spin chain is the $\widehat{\sgl}(2|1)_1$ WZNW model, that is the $\gl(2|1)$ GN model at zero coupling.
Finally, we mention the similarity with the situation for $\osp(R|2S)$ GN models~\cite{Bassi:1999ua, Saleur:2001cw, Saleur:2009bf}. 
Therefore, we conjecture that the continuous limit of the $\gl(n+N|N)$ spin chain $(V\otimes V^*)^{\otimes L}$ is the $\gl(n+N|N)$ GN model.

%%%%%%%%%%%%%%%%%%%%%%%%%%%%%%%%%%%%%%%%%%%
\subsection{Degree one excitation}
%%%%%%%%%%%%%%%%%%%%%%%%%%%%%%%%%%%%%%%%%%%

In this section we investigate degree 1 excitations of the $\gl(2m|1)$ spin chain which are of the type
~\eqref{eq:conj_1} and tend to the form~\eqref{eq:ideal_sol_1} in the thermodynamic limit.
As explained in sec.~\ref{sec:deg1}, the lowest lying excitation of degree 1 is precisely of this type. It corresponds to the antisymmetric tensor $t_0(1^m,1^m)$ and its root numbers are represented in fig.~\ref{fig:deg1}.

In the thermodynamic limit the roots
\begin{equation}\label{eq:identification} \{x_j^{(k)}\}^{\nu^{(k)}}_{j=1}=\{x_j^{(2m+1-k)}\}^{\nu^{(2m+1-k)}}_{j=1}\ , \quad k=1,\dots,m-1
\end{equation}
and the centers $\{x_j^{(\pm)}\}_{j=1}^{N_\pm}$, of the $\pm$-strange strings become dense on the real line.
We now derive the Lieb equations satisfied by their densities.
Collecting all BAE, in the grading $[+^m-+^m]$, we get
\begin{align}\label{eq:BAE_deg_1}
 -1&=e_{1}(x-\Lambda/2)^{L/2}e_{1}(x+\Lambda/2)^{L/2}E_2^{(1)}(x) E_{-1}^{(2)}(x)\ ,& x&\in\{x_j^{(1)}\}_{j=1}^{\nu^{(1)}}\\\notag
-1 &= E_{-1}^{(k-1)}(x)E_{2}^{(k)}(x) E_{-1}^{(k+1)}(x)\ ,\quad k=2,\dots,m-2\ ,& x&\in \{x_j^{(k)}\}_{j=1}^{\nu^{(k)}}\\\notag
-1&=E_{-1}^{(m-2)}(x)^2E_2^{(m-1)}(x)^2E_{-\frac{3}{2}}^{(+)}(x)E_{-\frac{1}{2}}^{(+)}(x)E_{-\frac{3}{2}}^{(-)}(x)E_{-\frac{1}{2}}^{(-)}(x)\ ,& x&\in \{x_j^{(m-1)}\}_{j=1}^{\nu^{(m-1)}}\\
+1&=E_{-\frac{3}{2}}^{(m-1)}(x)E_{-\frac{1}{2}}^{(m-1)}(x)E_2^{(\pm)}(x)E_1^{(\mp)}(x)^2\ ,& x&\in\{x_j^{(\pm)}\}_{j=1}^{N_\pm}\notag
\end{align}
where $E_t^{(k)}(x)=\prod_{i=1}^{\nu^{(k)}}e_t(x-x^{(k)}_j)$, for $k=1,\dots,m-1,\pm$.
The first two sets of BAE equations is just a rewriting of eqs.~\eqref{eq:BAE_short}. To get the last three we have multiplied the equations for same root appearing symmetrically in the Dynkin diagram~\ref{fig:deg1} at positions $m-1$ and $m+2$ or $m$ and $m+1$. 
Notice the $+$ sign appearing in the last two sets of equations.

Define the densities of roots $\rho_k(x^{(k)}_j):=\lim_{L\to \infty}1/L(x^{(k)}_{j+1}-x^{(k)}_j)$ and holes $\rho_k^h(x):=\sum_{j=1}^{n^{(k)}}\delta(x-\xi^{(k)}_j)/L$ for
$k=1,\dots,m-1,\pm$.
Taking the logarithm of eqs.~\eqref{eq:BAE_deg_1} and derivating w.r.t. the spectral parameters we get the Lieb equations
\begin{align}\label{eq:lieb_deg1}
 \tfrac{1}{2}a_1*
 (\delta_{\frac{\Lambda}{2}}+ \delta_{-\frac{\Lambda}{2}})&=\rho_1^h+\rho_1+a_2*\rho_1-a_1*\rho_2\\ \notag
0&=\rho_k^h +\rho_k +a_2* \rho_k - a_1*(\rho_{k-1}+\rho_{k+1})\ ,\quad k=2,\dots,m-2\\ \notag
0&=\rho_{m-1}^h + \rho_{m-1}+a_2*\rho_{m-1}-a_1*\rho_{m-2}-\tfrac{1}{2}(a_{\frac{3}{2}}+a_{\frac{1}{2}})*(\rho_++\rho_-)\\ \notag
0&=\rho^h_{\pm} + \rho_{\pm} +a_2* \rho_\pm +2 a_1* \rho_\mp - (a_{\frac{3}{2}}+a_{\frac{1}{2}})* \rho_{m-1}\ ,
\end{align}
where $\delta_x$ is the Dirac distribution centered at $x$, $a_t(x)=\tfrac{1}{2\pi}\tfrac{d\theta_t(x)}{dx}=\tfrac{1}{\pi}\tfrac{t/2}{x^2+t^2/4}$ and $*$ denotes the convolution $(f*g)(x)=\int_{-\infty}^{+\infty}dy\,f(x-y)g(y)$.
From eq.~\eqref{eq:spec_ham_mom_stag}, the energy and momentum of a root configuration~\eqref{eq:lieb_deg1} is 
\begin{align*}
 E&=-2\pi L\int_{-\infty}^{\infty} dx\, \rho_1(x)[a_1(x+\Lambda/2)+a_1(x-\Lambda/2)]\\
P&=L\int_{-\infty}^\infty dx\,\rho_1(x)[\theta_1(x+\Lambda/2)+\theta_1(x-\Lambda/2)]
\end{align*}

To solve for the root densities in terms of the hole densities it is useful to define the distributions
\begin{align*}
\rho_s&=\frac{\rho_+ +\rho_-}{2}\ ,& \rho_a&=\frac{\rho_+ -\rho_-}{2}\\
\rho^h_s&=\frac{\rho^h_+ +\rho^h_-}{2}\ ,& \rho^h_a&=\frac{\rho^h_+ -\rho^h_-}{2}\ .
\end{align*}
Notice that the distributions $\rho^h_a$, $\rho_a$ split from the rest of eqs.~\eqref{eq:lieb_deg1}
\begin{equation}\label{eq:sing_anti}
 0=\rho_a^h +\rho_a+a_2*\rho_a -2a_1*\rho_a\ ,
\end{equation}
while the Lieb equations for the remaining densities $\rho_1,\dots,\rho_{m-1},\rho_s$ look exactly like the Lieb equations for the staggered $\so(2m+1)$ fundamental  chain of length $L$, e.g.~\cite{Martins:1990ca}. The resolvent for the latter can be found in~\cite{ORW}.
Computing the solution of eqs.~\eqref{eq:BAE_deg_1} with no holes one gets
\begin{align}\label{eq:no_holes_deg1}
\rho^0_k(x)&=\frac{2}{n}\frac{\sin\tfrac{\pi k}{n}\cosh\pi\tfrac{2x+\Lambda}{n}}{\cosh\tfrac{2\pi(2x+\Lambda)}{n} -\cos \tfrac{2\pi k}{n}}+\frac{2}{n}\frac{\sin\tfrac{\pi k}{n}\cosh\pi\tfrac{2x-\Lambda}{n}}{\cosh\tfrac{2\pi(2x-\Lambda)}{n} -\cos \tfrac{2\pi k}{n}}\ ,\\
\rho^0_s(x)&=\frac{1}{2n}\frac{1}{\cosh \tfrac{\pi(2x+\Lambda)}{n}}+
\frac{1}{2n}\frac{1}{\cosh \tfrac{\pi(2x-\Lambda)}{n}}\ , \quad k=1,\dots,m-1\ ,\notag
\end{align}
where $n=2m-1$.
We want to stress that this is a \emph{formal} solution that does not correspond to any state in the spin chain. This is because the degree 1 excitations we are considering always contain holes.
However, it is  important to realize that the energy of the formal solution~\eqref{eq:no_holes_deg1} coincides with the actual vacuum energy~\eqref{eq:vac_en}.  We have already explained in sec.~\ref{sec:vac_en} why this happens.
As usual, the energy and momenta of holes can be expressed in terms of hole less densities
\begin{align*}
\epsilon_k (\xi) &= 4\pi \rho_k^0(\xi)\ ,& \frac{d p_k(\xi)}{d\xi}&=2\pi \rho_k^0(\xi)\\
\epsilon_\pm (\xi)&=2\pi \rho^0_s(\xi)\ ,& \frac{d p_\pm(\xi)}{d\xi}&=\pi \rho_k^0(\xi)
\end{align*}
Restricting to low lying states $|\xi|\ll \Lambda$ one gets the expected relativistic dispersion relations
\begin{align}\label{eq:old_particles}
\epsilon_k(\theta) &= 2c a m_k \cosh\theta \ ,& p_k(\theta) & = 2a m_k \sinh \theta \\
\epsilon_\pm(\theta) &= c a m \cosh\theta \ ,& p_\pm(\theta) & = a m \sinh \theta \label{eq:new_particles}
\end{align}
with the $\gl(n)$ Fermi velocity $c=4\pi/n$ and physical rapidity $\theta = 2\pi \xi/n$.
The masses are the same as in eqs.~(\ref{eq:cont_limit}, \ref{eq:bos_particles}).

It is certainly reassuring that degree 1 excitations have the same mass scale $m$ as degree 0 excitations. The doubling of masses~\eqref{eq:old_particles} w.r.t. the masses of particles in the bosonic lift can be understood as follows.
We have observed in sec.~\ref{sec:low_lying_deg1} that the imaginary parts of the odd roots tend to vanish for large $L$ and, therefore, we have neglected them in the thermodynamic limit.
However, due to the conjugation symmetry \eqref{eq:conj_1} of the solutions, this approximation leads to an identification~\eqref{eq:identification} of \emph{distinct} roots.
Therefore, the positions of holes in the distribution of even roots $\{x_j^{(k)}\}_{j=1}^{\nu^{(k)}}$ and $\{x_j^{(2m+1-k)}\}_{j=1}^{\nu^{(2m+1-k)}}$ have also been identified.
So, the excitation~\eqref{eq:old_particles} is in fact a state of two particles $1^k$ and $\bar{1}^k$ each of mass $m_k$ and rapidity $\theta$.
These are the old particles from the bosonic lift.
On the other hand, the particles~\eqref{eq:new_particles} are new.

The artificial binding of holes~\eqref{eq:old_particles} induced by the thermodynamic limit does not allow to derive meaningful $S$-matrices from eqs.~\eqref{eq:lieb_deg1}
Most probably, to correct the approach and recover the degrees of freedom lost in the thermodynamic limit one needs to distinguish between the two types of roots $\Img x_j^{(k)}>0$ and $\Img x_j^{(k)} < 0$, exactly as we did for the strange $\pm$-strings~\eqref{eq:ideal_sol_1}.
However, a finite volume treatment will be required, because the imaginary parts of even roots vanish in the thermodynamic limit.

The need for a finite volume approach can also be seen from eq.~\eqref{eq:sing_anti}.
Solving for the Fourier transform of $\hat{\rho}_a(p)=\int dp\,\exp(-2\pi i p x)\rho_a(x)$ one gets
\begin{equation}\label{eq:dist_anti_four}
\hat{\rho}_a(p\mid\xi^{(+)},\xi^{(-)})=\frac{i}{2L}e^{\pi |p|-\pi i p (\xi^{(+)}+\xi^{(-)})}\frac{\sin \pi p (\xi^{(+)}-\xi^{(-)})}{\sinh^2 \tfrac{\pi p }{2}}\ ,
\end{equation}
where we have considered only a single pair of $\pm$-holes.
This is no loss of generality, because from~\eqref{eq:lieb_deg1} the numbers of $\pm$-holes is always equal
\begin{equation*}
n^{\pm}=2(\nu^{(m-1)}-N_+-N_-) = 2(\nu^{(m-1)}-\nu^{(m)})\ .
\end{equation*}
and, therefore, they always come in pairs.
The Fourier transform~\eqref{eq:dist_anti_four} is singular and clearly must be regularized, because one has to satisfy the constraint
\begin{equation}\label{eq:charge_finally_found}
b:=\lim_{L\to \infty} \frac{N_+-N_-}{2L} = L \int_{-\infty}^{\infty} dx\, \rho_a(x)	=\hat{\rho}_a(0)\ ,
\end{equation}
where $b$ is a fixed real number parametrizing the state.
On a lattice of length $L$ the the momentum can take a minimal value of $p_\sim 1/L$. Using this value as a regulator one gets for a pair of $\pm$-holes from eqs.~(\ref{eq:dist_anti_four} \ref{eq:charge_finally_found}) 
\begin{equation*}
\xi^{(+)}-\xi^{(-)}=\frac{\pi b}{2i}\ ,
\end{equation*}
which clearly shows that for $b\neq 0$ the deviations of strange $\pm$-strings from the form~\eqref{eq:ideal_sol_1} of the solution in the thermodynamic limit have to be taken into account.
Notice that the energy does not depend on the continuous parameter~\eqref{eq:charge_finally_found} parametrizing the state.
Therefore, it is tempting to conclude that there is a \emph{continuum} of new particles~\eqref{eq:new_particles}.

%%%%%%%%%%%%%%%%%%%%%%%%%%%%%%%%%%%%%%
%%%%%%%%%%%%%%%%%%%%%%%%%%%%%%%%%%%%%%

\section{Conclusions and Outlook}

%%%%%%%%%%%%%%%%%%%%%%%%%%%%%%%%%%%%%%
%%%%%%%%%%%%%%%%%%%%%%%%%%%%%%%%%%%%%%

We have put on firm grounds the relationship between $\gl(n+N|N)$ integrable spin chains with $n$ fixed.
This allowed us to prove that all $\gl(n+N|N)$ spin chains $(V\otimes V^*)^{\otimes L}$ with $n,N>0$ possess in the continuum limit $2n-2$ multiplets of massive particles which scatter with $\gl(n)$ Gross-Neveu like $S$-matrices, namely their eigenvalues do not depend on $N$.
We concluded that the continuum theory is the $\gl(M|N)$ Gross-Neveu model.
Evidence that the massive spectrum is much richer, possibly continuous, was established on the example of $\gl(2m|1)$ chain.
Finally, our analysis of the thermodynamic limit strongly suggests that understanding the nature of new particles requires a finite volume treatment.

The question that begs the quickest answer is how to close the fusion
of $S$-matrices~\eqref{eq:smat_11} of the $\gl(M|N)$ Gross-Neveu model starting with just the vector multiplet and its antiparticles.

The $\gl(N|N)$ spin chains require a separate treatment, which we hope to report on later.\\

\noindent\textbf{Acknowledgements.} I would like to greatly thank Hubert Saleur, Volker Schomerus, Sergei Lukyanov and J\"org Techner for helpful discussions, important guidance and critical feedback. I am also grateful to Fabian  E{\ss}ler, Holger Frahm and Nikolay Gromov for sharing some of their expertise in solving numerically BAE.
The author thanks the Rutgers NHET center for their hospitality, where an important part of this paper was written, and SFB676 for partial financial support.  

\appendix

%%%%%%%%%%%%%%%%%%%%%%%%%%%%%%%%%%%%%%%%%%%%%%%%%%%%%%%%
%%%%%%%%%%%%%%%%%%%%%%%%%%%%%%%%%%%%%%%%%%%%%%%%%%%%%%%%

\section{Highest weight vectors}
\label{sec:proofs}

%%%%%%%%%%%%%%%%%%%%%%%%%%%%%%%%%%%%%%%%%%%%%%%%%%%%%%%%
%%%%%%%%%%%%%%%%%%%%%%%%%%%%%%%%%%%%%%%%%%%%%%%%%%%%%%%%

Let $T$ be the indecomposable direct summand of $t(\lambda,\mu)$ with traceless submodule $t_0(\lambda,\mu)$.
Without loss of generality, one can assume that $(\lambda,\mu)$ is of admissible shape in the sense of eq.~\ref{eq:restr_form}.
The filtration~\eqref{eq:filtr} implies a similar filtration for $T$
\begin{equation*}
t_0(\lambda,\mu)=T_0\subset T_1\subset \cdots \subset T_f=T 
%\label{eq:}
\end{equation*}
where we have defined the submodules $T_n=T\cap t_n(\lambda,\mu)$.
Again as for $t(\lambda,\mu)$ one has
\begin{equation}
T_n/T_{n-1}\simeq \bigoplus_{(\lambda',\mu')\in D_n} t_0(\lambda',\mu')\ ,
\label{eq:tr_sq}
\end{equation}
where the elements $(\lambda',\mu')$ of $D_n$ must be pairs of diagrams with $f-n$ boxes such that $\lambda'\subset \lambda$ and $\mu'\subset \mu$.
In particular,  they are also admissible. According to assumption~\ref{ass:hw}, every direct summand $t_0(\lambda',\mu')$ in eq.~\eqref{eq:tr_sq} can be generated from a highest weight vector $\mathbf{v}(\lambda',\mu')/T_{n-1}\in t_0(\lambda',\mu')$ w.r.t. some Borel subalgebra $\mathfrak{b}_\Sigma$, where $\mathbf{v}(\lambda',\mu')\in T_n$.
Assume that $\mathbf{v}(\lambda',\mu')$ is a highest weight vector w.r.t. $\mathfrak{b}_\Sigma$.
Then $\mathbf{v}(\lambda',\mu')$ cannot generate more then a Kac module within $T_n$, which according to assumption~\ref{ass:hw} is precisely $t_0(\lambda',\mu')$.
Thus, $t_0(\lambda',\mu')$ is a submodule of $T_n$ and therefore also of $T$. This completes the proof of claim~\ref{claim:cv}.

Notice that if  $T(\lambda,\mu)$ is of the usual ``diamond'' form, that is  its socle and top are irreducible, then the existence of two submodules $t_0(\lambda,\mu),t_0(\lambda',\mu')\subset T(\lambda,\mu)$ contradicts the simplicity of the socle and, therefore, all highest weight vectors of $T(\lambda,\mu)$ belong to $t_0(\lambda,\mu)$.

%%%%%%%%%%%%%%%%%%%%%%%%%%%%%%%%%%%%%%%%%%%%%%%%%%%%%%%%
%%%%%%%%%%%%%%%%%%%%%%%%%%%%%%%%%%%%%%%%%%%%%%%%%%%%%%%%

\section{Cohomological reduction of Kac modules}
\label{sec:B}

%%%%%%%%%%%%%%%%%%%%%%%%%%%%%%%%%%%%%%%%%%%%%%%%%%%%%%%%
%%%%%%%%%%%%%%%%%%%%%%%%%%%%%%%%%%%%%%%%%%%%%%%%%%%%%%%%

In this section we prove the claim~\ref{claim:coh} of sec.~\ref{sec:rl} and adopt all the notations leading to it.

Assume that we have a solution for $p<P$ and a corresponding Bethe vector $\omega_1$. It comes together with a solution for $p=P$ and a corresponding Bethe vector $\omega_2$.
The transfer matrices~\eqref{eq:tmat} yield the same eigenvalues on $\omega_1$ and $\omega_2$. Therefore, they must be part of a reducible indecomposable module or, more precisely due to corollary~1 of sec.~\ref{sec:tens}, part of a Kac submodule of the latter.
Moreover, because $\wt(\omega_2)<\wt(\omega_1)$ the module generated by $\omega_2$ must be a submodule of the module generated by $\omega_1$
\begin{equation}\label{eq:int_bsubalg}
 \mathcal{U}(\gl(M|N)) \cdot \omega_2\subset \mathcal{U}(\gl(M|N))\cdot \omega_1 \subseteq K_{\mathfrak{b}}(\Lambda). 
\end{equation}
Here $\Lambda$ is the highest weight of the Kac module w.r.t. some Borel subalgebra $\mathfrak{b}$.
Obviously, the highest weight must be atypical $\langle \Lambda,\alpha_k\rangle=0$ in order for \eqref{eq:int_bsubalg} to hold.
Recall that the Borel subalgebras $\mathfrak{b}_\Sigma$ and $\mathfrak{b}$ of $\gl(M|N)$ define each a $\mathbb{Z}$--graded decomposition of the latter~\cite{Frappat:1996pb}
\begin{align*}
\gl(M|N)&\simeq \mathfrak{g}^{-1}_{\phantom{\Sigma}}\oplus \mathfrak{g}^0_{\phantom{\Sigma}}\oplus\mathfrak{g}^1_{\phantom{\Sigma}}\ , \qquad  \mathfrak{g}^1_{\phantom{\Sigma}}  \subset \mathfrak{b}_{\phantom{\Sigma}}\\
 \gl(M|N)&\simeq \mathfrak{g}^{-1}_\Sigma \oplus \mathfrak{g}^0_\Sigma\oplus\mathfrak{g}^1_\Sigma\ , \qquad \mathfrak{g}^1_\Sigma  \subset \mathfrak{b}_\Sigma   \ .
\end{align*}
Recall the definition of Kac modules
\begin{equation}\label{eq:Kac_mod}
K_{\mathfrak{b}}(\Lambda)\simeq \bigwedge \left(\mathfrak{g}^{-1}\right)\otimes_{\mathcal{U}(\mathfrak{g}^0\oplus\mathfrak{g}^1)}S_0(\Lambda)\ ,
\end{equation}
where $S_0(\Lambda)$ is an irreducible representation of $\mathfrak{g}_0$ of highest weight $\Lambda$ trivially extended to a representation of $\mathfrak{g}^0\oplus\mathfrak{g}^1$.
Then, due to the embeddings~\eqref{eq:int_bsubalg}, one must have $Q\in \mathfrak{g}^{-1}\cap \mathfrak{g}^{-1}_\Sigma$.
Notice now that Kac modules have zero superdimension. This is a necessary condition for vanishing $Q$--cohomology. Indeed, given $Q\in\mathfrak{g}^{-1}$ it is straightforward to show that the $Q$--cohomology of the Kac module defined as in~\eqref{eq:Kac_mod} vanishes.
For more details, see the proof of \cite{Candu:2010yg} for the vanishing of the cohomology of projective modules.

Notice that while projective modules have zero cohomology w.r.t. any choice of $Q$, the Kac module~\eqref{eq:Kac_mod}, on the other hand, has zero cohomology only w.r.t. $Q\in\mathfrak{g}^{-1}$.

%%%%%%%%%%%%%%%%%%%%%%%%%%%%%%%%%%%%%%%%%%%%%%%%%%%%%%%%
%%%%%%%%%%%%%%%%%%%%%%%%%%%%%%%%%%%%%%%%%%%%%%%%%%%%%%%%
%%%%%%%%%%%%%%%%%%%%%%%%%%%%%%%%%%%%%%%%%%%%%%%%%%%%%%%%

\bibliographystyle{hep}
\bibliography{bae}

\newcommand{\etalchar}[1]{$^{#1}$}
\begin{thebibliography}{CMQ{\etalchar{+}}10}

\bibitem[AL79]{Andrei:1979sq}
N.~Andrei and J.~H. Lowenstein, \textsl{ {Diagonalization of the Chiral
  Invariant Gross-Neveu Hamiltonian}},
\newblock Phys. Rev. Lett. \textbf{ 43}, 1698 (1979).

\bibitem[AL80a]{Andrei:1979un}
N.~Andrei and J.~H. Lowenstein, \textsl{ {A Direct Calculation of the S Matrix
  of the Chiral Invariant Gross-Neveu Model}},
\newblock Phys. Lett. \textbf{ B91}, 401 (1980).

\bibitem[AL80b]{Andrei:1979wy}
N.~Andrei and J.~H. Lowenstein, \textsl{ {Derivation of the Chiral Gross-Neveu
  Spectrum for Arbitrary SU(N) Symmetry}},
\newblock Phys. Lett. \textbf{ B90}, 106 (1980).

\bibitem[BCH{\etalchar{+}}94]{staircase}
G.~Benkart, M.~Chakrabarti, T.~Halverson, R.~Leduc, C.~Lee and J.~Stroomer,
  \textsl{ {Tensor product representations of general linear groups and their
  connections with Brauer algebras}},
\newblock J. Algebra \textbf{ 166}, 529 (1994).

\bibitem[{Ber}]{Bernard:1995as}
D.~{Bernard}, \textsl{ {(Perturbed) Conformal Field Theory Applied To 2D
  Disordered Systems: An Introduction}},
\newblock {hep-th/9509137}.

\bibitem[BL00]{Bassi:1999ua}
Z.~S. Bassi and A.~LeClair, \textsl{ {The exact S-matrix for an $\osp(2|2)$
  disordered system}},
\newblock Nucl. Phys. \textbf{ B578}, 577 (2000), {hep-th/9911105}.

\bibitem[BR08]{Belliard:2008di}
S.~Belliard and E.~Ragoucy, \textsl{ {Nested Bethe ansatz for 'all' closed spin
  chains}},
\newblock J.Phys.A \textbf{ A41}, 295202 (2008), {0804.2822}.

\bibitem[Car86]{Cardy:1986ie}
J.~L. Cardy, \textsl{ {Operator Content of Two-Dimensional Conformally
  Invariant Theories}},
\newblock Nucl. Phys. \textbf{ B270}, 186 (1986).

\bibitem[CCMS10]{Candu:2010yg}
C.~Candu, T.~Creutzig, V.~Mitev and V.~Schomerus, \textsl{ {Cohomological
  Reduction of Sigma Models}},
\newblock JHEP \textbf{ 05}, 047 (2010), {1001.1344}.

\bibitem[CMQ{\etalchar{+}}10]{Candu:2009ep}
C.~Candu, V.~Mitev, T.~Quella, H.~Saleur and V.~Schomerus, \textsl{ {The Sigma
  Model on Complex Projective Superspaces}},
\newblock JHEP \textbf{ 02}, 015 (2010), {0908.0878}.

\bibitem[CR09]{Creutzig:2008an}
T.~Creutzig and P.~B. Ronne, \textsl{ {The $\GL(1|1)$-symplectic fermion
  correspondence}},
\newblock Nucl. Phys. \textbf{ B815}, 95 (2009), {0812.2835}.

\bibitem[CS09a]{Candu:2008vw}
C.~Candu and H.~Saleur, \textsl{ {A lattice approach to the conformal
  $\OSp(2S+2|2S)$ supercoset sigma model. Part I: Algebraic structures in the
  spin chain. The Brauer algebra}},
\newblock Nucl. Phys. \textbf{ B808}, 441 (2009), {0801.0430}.

\bibitem[CS09b]{Candu:2008yw}
C.~Candu and H.~Saleur, \textsl{ {A lattice approach to the conformal
  $\OSp(2S+2|2S)$ supercoset sigma model. Part II: The boundary spectrum}},
\newblock Nucl. Phys. \textbf{ B808}, 487 (2009), {0801.0444}.

\bibitem[CVDM08]{martin}
A.~Cox, M.~D. Visscher, S.~Doty and P.~Martin, \textsl{ {On the blocks of the
  walled Brauer algebra}},
\newblock J. Algebra \textbf{ 320}, 169 (2008).

\bibitem[DdV89]{Destri:1987ug}
C.~Destri and H.~J. de~Vega, \textsl{ {Light Cone Lattices and the Exact
  Solution of the Chiral Fermion and Sigma Models}},
\newblock J. Phys. \textbf{ A22}, 1329 (1989).

\bibitem[DN98]{Doikou:1998xi}
A.~Doikou and R.~I. Nepomechie, \textsl{ {Bulk and boundary S matrices for the
  SU(N) chain}},
\newblock Nucl. Phys. \textbf{ B521}, 547 (1998), {hep-th/9803118}.

\bibitem[EFS05]{Essler:2005ag}
F.~H.~L. E{\ss}ler, H.~Frahm and H.~Saleur, \textsl{ {Continuum Limit of the
  Integrable $\sgl(2|1)$ $3-\bar{3}$ Superspin Chain}},
\newblock Nucl. Phys. \textbf{ B712}, 513 (2005), {cond-mat/0501197}.

\bibitem[EK94]{1994IJMPB...8.3243E}
F.~H.~L. {E{\ss}ler} and V.~E. {Korepin}, \textsl{ {Spectrum of Low-Lying
  Excitations in a Supersymmetric Extended Hubbard Model}},
\newblock International Journal of Modern Physics B \textbf{ 8}, 3243 (1994),
  {cond-mat/9307019}.

\bibitem[Fad96]{Faddeev:1996iy}
L.~D. Faddeev, \textsl{ {How Algebraic Bethe Ansatz works for integrable
  model}},
\newblock (1996), {hep-th/9605187}.

\bibitem[FSS96]{Frappat:1996pb}
L.~Frappat, P.~Sorba and A.~Sciarrino, \textsl{ {Dictionary on Lie
  superalgebras}},
\newblock (1996), {hep-th/9607161}.

\bibitem[GCA98]{ram}
B.~G., S.~C.L. and R.~A., \textsl{ {Tensor product representations for
  orthosymplectic Lie superalgebras}},
\newblock J. Pure Appl. Algebra \textbf{ 130}, 1 (1998).

\bibitem[GLL00]{2000NuPhB.583..475G}
S.~{Guruswamy}, A.~{LeClair} and A.~W.~W. {Ludwig}, \textsl{ {$\gl(N|N)$
  Super-current algebras for disordered Dirac fermions in two dimensions}},
\newblock Nuclear Physics B \textbf{ 583}, {475} (2000), {cond-mat/9909143}.

\bibitem[IJS08]{2006cond.mat.12037I}
Y.~Ikhlef, J.~Jacobsen and H.~Saleur, \textsl{ {A staggered six-vertex model
  with non-compact continuum limit}},
\newblock Nuclear Physics B \textbf{ 789}(3), 483 (2008), {cond-mat/0612037}.

\bibitem[Kac77a]{Kac77b}
V.~G. Kac, \textsl{ {Characters of typical representations of classical Lie
  superalgerbas}},
\newblock Comm. Alg. \textbf{ 5}, 889 (1977).

\bibitem[Kac77b]{Kac77a}
V.~G. Kac, \textsl{ {Lie Superalgebras}},
\newblock Adv. Math. \textbf{ 26}, 8 (1977).

\bibitem[KKS79]{PhysRevD.20.897}
R.~K\"oberle, V.~Kurak and J.~A. Swieca, \textsl{ {Scattering theory and 1/N
  expansion in the chiral Gross-Neveu model}},
\newblock Phys. Rev. D \textbf{ 20}, 897 (1979).

\bibitem[KP91]{pearce}
A.~Kl\"umper and P.~A. Pearce, \textsl{ {Analytic calculation of scaling
  dimensions: Tricritical hard squares and critical hard hexagons}},
\newblock J. Stat. Phys. \textbf{ 64}, 13 (1991).

\bibitem[LeC09]{LeClair:2007aj}
A.~LeClair, \textsl{ {The gl$(1|1)$ super-current algebra: the role of twist
  and logarithmic fields}},
\newblock Adv. Theor. Math. Phys. \textbf{ 13}, 259 (2009), {0710.2906}.

\bibitem[Mar91]{Martins:1990ca}
M.~J. Martins, \textsl{ {Fractional strings hypothesis and nonsimple laced
  integrable models}},
\newblock J. Phys. \textbf{ A24}, L159 (1991).

\bibitem[Mar95]{Martins:1995bb}
M.~J. Martins, \textsl{ {Bethe ansatz solution of the OSP$(1|2n)$ invariant
  spin chain}},
\newblock Phys. Lett. \textbf{ B359}, 334 (1995), {hep-th/9502135}.

\bibitem[ORW87]{ORW}
E.~Ogievetsky, N.~Reshetikhin and P.~Wiegmann, \textsl{ {The Principal Chiral
  Field in Two-Dimensions on Classical Lie Algebras: The Bethe Ansatz Solution
  and Factorized Theory of Scattering}},
\newblock Nucl. Phys. \textbf{ B280}, 45 (1987).

\bibitem[OW86]{Ogievetsky:1986hu}
E.~Ogievetsky and P.~Wiegmann, \textsl{ {Factorized S Matrix and the Bethe
  Ansatz for Simple Lie Groups}},
\newblock Phys. Lett. \textbf{ B168}, 360 (1986).

\bibitem[Sal00]{Saleur:1999cx}
H.~Saleur, \textsl{ {The continuum limit of sl$(N|K)$ integrable super spin
  chains}},
\newblock Nucl. Phys. \textbf{ B578}, 552 (2000), {solv-int/9905007}.

\bibitem[Ser01]{Sergeev}
A.~Sergeev, \textsl{ {An analog of the classical invariant theory for Lie
  superalgebras I, II}},
\newblock Michigan Math. J. \textbf{ 49}, 113, 147 (2001).

\bibitem[SP10]{Saleur:2009bf}
H.~Saleur and B.~Pozsgay, \textsl{ {Scattering and duality in the 2 dimensional
  $\OSP(2|2)$ Gross Neveu and sigma models}},
\newblock JHEP \textbf{ 02}, 008 (2010), {0910.0637}.

\bibitem[SS07]{Saleur:2006tf}
H.~Saleur and V.~Schomerus, \textsl{ {On the $\SU(2|1)$ WZNW model and its
  statistical mechanics applications}},
\newblock Nucl. Phys. \textbf{ B775}, 312 (2007), {hep-th/0611147}.

\bibitem[Sut75]{Sutherland:1975vr}
B.~Sutherland, \textsl{ {A General Model for Multicomponent Quantum Systems}},
\newblock Phys. Rev. \textbf{ B12}, 3795 (1975).

\bibitem[SWK00]{Saleur:2000bq}
H.~Saleur and B.~Wehefritz-Kaufmann, \textsl{ {Thermodynamics of the complex
  su(3) Toda theory}},
\newblock Phys. Lett. \textbf{ B481}, 419 (2000), {hep-th/0003217}.

\bibitem[SWK02]{Saleur:2001cw}
H.~Saleur and B.~Wehefritz-Kaufmann, \textsl{ {Integrable quantum field
  theories with OSP$(m|2n)$ symmetries}},
\newblock Nucl. Phys. \textbf{ B628}, 407 (2002), {hep-th/0112095}.

\bibitem[Tsu98]{Tsuboi:1998ne}
Z.~Tsuboi, \textsl{ {Analytic Bethe Ansatz And Functional Equations Associated
  With Any Simple Root Systems Of The Lie Superalgebra $\sgl(r+1|s+1)$}},
\newblock Physica \textbf{ A252}, 565 (1998).

\bibitem[Vla86]{vlad}
A.~A. Vladimirov, \textsl{ {Proof of the invariance of the Bethe-ansatz
  solutions under complex conjugation}},
\newblock Theoretical and Mathematical Physics \textbf{ 66}, 102 (1986).

\bibitem[Wey53]{Weyl}
H.~Weyl,
\newblock \textsl{ {The classical groups: their invariants and
  representations}},
\newblock Princeton, 1953.

\bibitem[ZZ79]{Zamolodchikov:1978xm}
A.~B. Zamolodchikov and A.~B. Zamolodchikov, \textsl{ {Factorized S-matrices in
  two dimensions as the exact solutions of certain relativistic quantum field
  models}},
\newblock Annals Phys. \textbf{ 120}, 253 (1979).

\end{thebibliography}

%%%%%%%%%%%%%%%%%%%%%%%%%%%%%%%%%%%%%%%%%%%%%%%%%%%%%%%%
%%%%%%%%%%%%%%%%%%%%%%%%%%%%%%%%%%%%%%%%%%%%%%%%%%%%%%%%
%%%%%%%%%%%%%%%%%%%%%%%%%%%%%%%%%%%%%%%%%%%%%%%%%%%%%%%%
%%%%%%%%%%%%%%%%%%%%%%%%%%%%%%%%%%%%%%%%%%%%%%%%%%%%%%%%

\end{document}